# Second-order homogenization of periodic materials based on asymptotic approximation of the strain energy: formulation and validity limits.

Andrea Bacigalupo[1]

Department of Civil, Chemical and Environmental Engineering,
University of Genoa, via Montallegro, 1, 16145 Genoa, Italy

**Abstract**

In this paper a second-order homogenization approach for periodic material is derived from an appropriate representation of the down-scaling that correlates the micro-displacement field to the macro-displacement field and the macro-strain tensors involving unknown perturbation functions. These functions take into account of the effects of the heterogeneities and are obtained by the solution of properly defined recursive cell problems. Moreover, the perturbation functions and therefore the micro-displacement fields result to be sufficiently regular to guarantee the anti-periodicity of the traction on the periodic unit cell. A generalization of the macro-homogeneity condition is obtained through an asymptotic expansion of the mean strain energy at the micro-scale in terms of the microstructural characteristic size $\varepsilon$; the obtained overall elastic moduli result to be not affected by the choice of periodic cell. The coupling between the macro- and micro-stress tensor in the periodic cell is deduced from an application of the generalised macro-homogeneity condition applied to a representative portion of the heterogeneous material (cluster of periodic cell). The correlation between the proposed asymptotic homogenization approach and the computational second-order homogenization methods (which are based on the so called *quadratic ansätze*) is obtained through an approximation of the macro-displacement field based on a second-order Taylor expansion. The form of the overall elastic moduli obtained through the two homogenization approaches, here proposed, is analyzed and the differences are highlighted.

---

[1] andrea.bacigalupo@unige.it

An evaluation of the developed method in comparison with other recently proposed in literature is carried out in the example where a three-phase orthotropic material is considered. The characteristic lengths of the second-order equivalent continuum are obtained by both the asymptotic and the computational procedures here analyzed. The reliability of the proposed approach is evaluated for the case of shear and extensional deformation of the considered two-dimensional infinite elastic medium subjected to periodic body forces; the results from the second-order model are compared with those of the heterogeneous continuum.

# 1 Introduction

The control of the static and dynamic properties of composite materials having periodic microstructure is a topic of continuing interest that motivates the use of multi-scale approaches. In fact, the solution of static and dynamic problems of heterogeneous materials is in general very labour intensive and may be applied only to simple materials like stratified materials. For this reason, in order to obtain a synthetic mechanical description of periodic materials it is necessary to consider different scales of observation that lead to homogenization techniques, mostly based on a classical description of the Cauchy both at the micro- and at the macro-scale. However, if the size of the microstructural components is not negligible with respect to the structural dimension, or to the elastic wavelength, or in the presence of high strain gradients, the standard approach based on first order homogenization may present some disadvantages. Non-local continua may provide better results by introducing characteristic lengths in the constitutive model which allow better descriptions of the material response both in presence of strong stress gradients and in case of the post-elastic response of brittle constituents. In case of periodic materials, described by a periodic cell, different non-local homogenization procedures have been proposed and applied, such as the *asymptotic* [1-8], the *variational-asymptotic methods* [9,10] and the *computational approaches* [11-19].

The asymptotic and variational-asymptotic techniques provide a mathematically rigorous tool of higher-order homogenization of periodic linear elastic heterogeneous materials. In these methods the kinematics at the micro- and macro-scale are coupled through an asymptotic expansion of the micro-displacement in terms of characteristic size $\varepsilon$ of the microstructure. This expansion depends both on the macro-strains and on unknown perturbation functions accounting for the effects of the heterogeneities. The perturbations are determined by solving non-homogeneous cell problems with body forces which depend on the geometrical and mechanical properties of the microstructure and with periodic boundary conditions prescribed on the cell. Dynamic homogenization approaches based on the asymptotic [20,21] and variational-asymptotic approaches [22,23] have been developed with the aim of considering higher order inertia terms within a constitutive classical homogenization. Other dynamic homogenization methods are based on the high-contrast [24-29] and high frequency [30,31] asymptotic homogenization techniques. These last procedures appear to be rather laborious and computationally burdensome even if they allow the description of frequency band-gaps.

The approaches based on the computational homogenization (computational approaches) of the periodic cell developed for Cosserat and micromorphic homogenization [11,12,19] and second-order homogenization [13-18] seem to be very attractive because they present a simple formulation and are computationally less burdensome in comparison to the asymptotic and variational-asymptotic homogenization techniques. In these approaches the micro-displacement field is expressed as the superposition of a macroscopic displacement field that is assumed in polynomial form in terms of the macro-strains and of an unknown micro-displacement fluctuation field. The latter is obtained through the solution of homogeneous cell problems with vanishing body forces and prescribed periodic boundary condition. In most of the second-order homogenization methods the micro-displacement fluctuation field is assumed periodic and not all the second-order strain components can be controlled by the inhomogeneous periodic boundary conditions so that suitable integral conditions are introduced [13-15,17]. Moreover, unlike the first-order homogenization, these approaches do not guarantee the continuity of the micro-displacement field across the interface between adjacent cells in case of prescribed homogeneous second-order strain (or intrinsic strain and higher order strain). A second-order homogenization model that preserves only the continuity of the displacement field has been developed in [18] in which a fluctuation field is assumed non-periodic and generalised periodic boundary conditions for the second-order homogenization are introduced. The critical analysis developed in [19] concerning the second-order homogenization models highlights their limitations, the relevant open issues and remarks on the need to introduce a perturbation field of the non-periodic displacement field (as assumed in [16,18]). Moreover, it is also pointed out how the macro-strain components cannot be directly controlled by the coefficients of the quadratic polynomial and thus the up-scaling has to be governed by relations that depend, in general, on the micro-displacement fluctuation field. In [19] is also emphasized how these fluctuation field may depend on the choice of the periodic cell and on its characteristic size. In this regard a computational strategy is introduced to unambiguously determine this micro-displacement fluctuation field. Notably, a sufficiently large cluster is considered in order to apply the boundary condition as remotely from a central unit cell. Recently a generalization of the Hill-Mandel lemma to the case of strain-gradient continuum in terms of strain energy densities is proposed in [8], together with a consistent definition of the up-scaling of the macro-stress and of the hyperstresses. This up-scaling has allowed to obtain the macroscopic balance equation of a multipolar continuum as averaged over the periodic cell of the microscopic balance equation through a suitable truncation of the average equation of infinite order [1].

The critical analysis of the second-order computational homogenization developed in [19] and the generalization of Hill-Mandel lemma in [8] for strain gradient continuum have motivated this paper, where a non-local homogenization technique, based on the asymptotic and variational-asymptotic approaches [1,9], is developed and a further generalization of the macro-homogeneity condition is obtained. The here proposed asymptotic approach (Section 4) preserves the simplicity of the computational approaches, provides a continuous micro-displacement field across the interfaces between adjacent periodic cells and ensures the anti-periodicity of the tractions on the boundary of the cell through an appropriate description of the down-scaling. The micro-displacement field is obtained by superimposing to macro-displacement an appropriate representation of the micro-displacement fluctuation field. This micro-displacement fluctuation field is assumed as the superposition of two unknown functions each of them associated to the first-order and to the second-order strain, respectively. The perturbation functions, which depend only on the properties of the microstructure, are obtained through the solution of elastic non-homogeneous problems on the cell with periodic boundary conditions and normalization condition. To eliminate the fast oscillations of the local variables (with a period smaller than the characteristic size of the microstructure) an averaging procedure is introduced, in accordance to [9], that is carried out with reference to all possible realizations obtained from a translation of the heterogeneous microstructure of the medium. Moreover, a generalization of the macro-homogeneity condition applied to a representative portion of the heterogeneous material (cluster of periodic cell) is proposed, wherein the *mean strain* *energy* of the cluster is set equal to that of the second-order continuum. Notably, the mean strain energy of the cluster is expressed through an asymptotic expansion in terms of the microstructural characteristic size $\varepsilon$ (that depends both on the perturbation functions and on the macro-strain variables) suitably truncated following the application of the divergence theorem. According to this macro-homogeneity condition, the elastic moduli of the equivalent homogeneous continuum are evaluated and in the second order overall elastic moduli is possible to include terms associated to the third-order strain tensor. Conversely, these additional terms may not be determined through an equality of the elastic energy density. Recalling that the second-order computational homogenization is based on the quadratic ansätze [19], correlations with the computational approaches are analysed (Section 5) approximating the macro-displacement with a second order Taylor expansion. The overall elastic moduli that are obtained by computational homogenization model proposed in Section 5 are compared with those obtained through the asymptotic homogenization model developed in Section 4. Notably, the study of the correlations between these two different approaches, considering also that the computational approach leads to similar results to those obtained through other approaches in [8] and [22], may be regarded as a purpose of this paper.

To assess the reliability of the homogenization techniques, a three-phase orthotropic composite having a quasi-stratified microstructure is considered in the example (Section 6). The second-order homogenization is carried out by a FE analysis of the unit cell and the characteristic lengths of the second-order equivalent continuum model are obtained and compared with those obtained by the homogenization methods proposed in [8] and [18,22]. Finally, with reference to the three-phase material, the shear and extensional problems of the heterogeneous two-dimensional medium subjected to periodic body forces along the orthotropic axes are analysed both as a heterogeneous classical continuum and as a homogenized second-order continuum. The corresponding results are compared and discussed in order to identify the validity limits of the proposed technique also in relation to the approaches developed in [8] and [18,22].

## 2 Multi-scale modelling of materials with periodic microstructure

Let us consider a two-dimensional heterogeneous body characterised by a periodic microstructure (figure 1.a) undergoing small displacements; its constituent are modelled as an elastic Cauchy continuum. A generic point of the heterogeneous model is identified by the vector position $\mathbf{x} = x_1\mathbf{e}_1 + x_2\mathbf{e}_2$ referred to a system of coordinates with origin at point O and orthogonal base $(\mathbf{e}_1, \mathbf{e}_2)$. A periodic cell $\mathcal{A} = [0,\varepsilon] \times [0,\delta\varepsilon]$ having size $\varepsilon$ is shown in figure 1.b, together with the periodicity vectors $\mathbf{v}_1$ and $\mathbf{v}_2$ referred for simplicity to the case of orthogonal vectors. Moreover, the elasticity tensor $\mathbb{C}^{m,\varepsilon}(\mathbf{x})$ results $\mathcal{A}$–periodic, i.e. $\mathbb{C}^{m,\varepsilon}(\mathbf{x} + \mathbf{v}_i) = \mathbb{C}^{m,\varepsilon}(\mathbf{x})$, $i = 1,2$, $\forall \mathbf{x} \in \mathcal{A}$. The heterogeneous medium is considered subjected to a system of $\mathcal{L}$-periodic body forces $\mathbf{f}$ with vanishing mean value on $\mathcal{L}$, where $\mathcal{L} = [0,L] \times [0,\delta L]$ and $L$ is a multiple of $\varepsilon$ (figure 1.a) so that the portion $\mathcal{L}$ may be assumed as representative of the overall body.

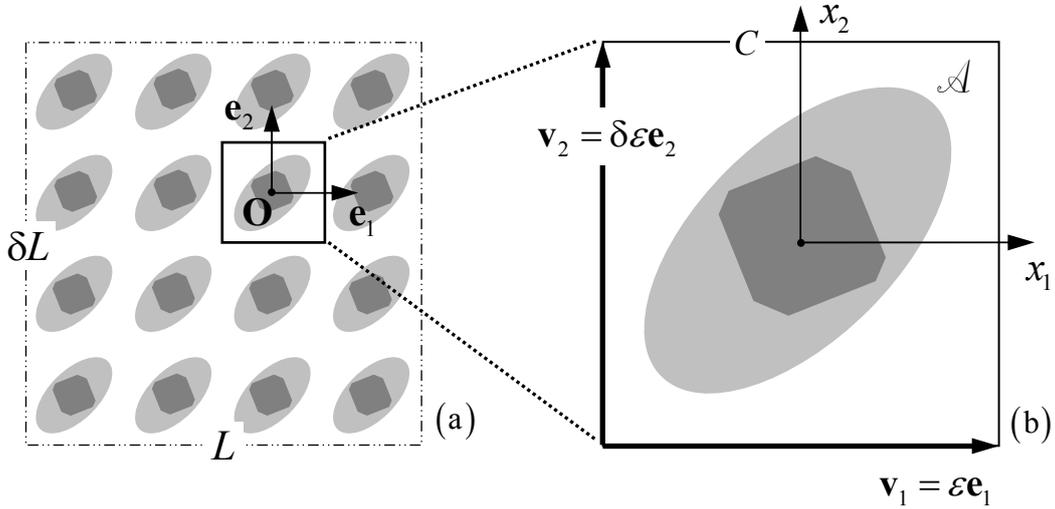

Figure 1: (a) Heterogeneous material – Periodic domain $\mathcal{L}$; (b) Periodic cell $\mathcal{A}$ and periodicity vectors.

The relevant micro-fields are the micro-displacement $\mathbf{u}(\mathbf{x})$, the micro-strain $\boldsymbol{\varepsilon}(\mathbf{x}) = sym\nabla\mathbf{u}(\mathbf{x})$ and the micro-stress $\boldsymbol{\sigma}(\mathbf{x}) = \mathbb{C}^{m,\varepsilon}(\mathbf{x})\boldsymbol{\varepsilon}(\mathbf{x})$, the latter which has to satisfy the local equilibrium equation $\nabla \cdot \boldsymbol{\sigma}(\mathbf{x}) + \mathbf{f}(\mathbf{x}) = \mathbf{0}$ in the domain $\mathcal{A}$. In accordance with conventional approaches of asymptotic homogenization [1,9] the local elasticity tensor is represented in the unit cell $Q = [0,1] \times [0,\delta]$ rescaling to $\mathbb{C}^{m,\varepsilon}(\mathbf{x}) = \mathbb{C}^m\left(\boldsymbol{\xi} = \dfrac{\mathbf{x}}{\varepsilon}\right)$, where $\boldsymbol{\xi} \in Q$ assumes the meaning of microscopic coordinate (or *fast variable*). The equilibrium equations in terms of the micro-displacement field result

$$\nabla \bullet \left( \mathbb{C}^m \left( \frac{\mathbf{x}}{\varepsilon} \right) \nabla \mathbf{u}(\mathbf{x}) \right) + \mathbf{f}(\mathbf{x}) = \mathbf{0}, \tag{1}$$

and allows to represent the micro-displacement in the form $\mathbf{u}\left( \mathbf{x}, \boldsymbol{\xi} = \dfrac{\mathbf{x}}{\varepsilon} \right)$, which results to be $\mathscr{L}$-periodic with respect to the variable $\mathbf{x}$ (see [1,9]). The solution of the elastic problem (1) is in general computationally cumbersome because the coefficients are $\mathscr{A}$-periodic. Therefore it is preferable to replace the heterogeneous system with a homogeneous equivalent one whose behaviour may be described either by a Cauchy continuum or by a non-local continuum (micromorphic [32] or multipolar [33]) through $\mathscr{L}$-periodic state variables that depend only of the macroscopic coordinate $\mathbf{x}$ (or *slow variable*). To this purpose an asymptotic expansion for the micro-displacement is considered in terms of the parameter $\varepsilon$ that keeps the dependence on the slow variable $\mathbf{x}$ separate from the fast one $\boldsymbol{\xi}$ (so that two distinct scale are represented); this procedure will be illustrated in Section 3.

In the second gradient continuum [32,33] the macro-displacement $\mathbf{U}(\mathbf{x})$ of component $U_i$ is defined at point $\mathbf{x}$ in the reference $(\mathbf{e}_i, i = 1, 2)$ together with the displacement gradient $\mathbf{H}(\mathbf{x}) = H_{ij} \mathbf{e}_i \otimes \mathbf{e}_j = \nabla \mathbf{U}(\mathbf{x}) = \dfrac{\partial U_i}{\partial x_j} \mathbf{e}_i \otimes \mathbf{e}_j$ and the second-order strain $\boldsymbol{\kappa}(\mathbf{x}) = \kappa_{ijk} \mathbf{e}_i \otimes \mathbf{e}_j \otimes \mathbf{e}_k = \nabla \otimes \nabla \mathbf{U}(\mathbf{x}) = \dfrac{\partial^2 U_i}{\partial x_j \partial x_k} \mathbf{e}_i \otimes \mathbf{e}_j \otimes \mathbf{e}_k$. As usual, the displacement gradient can be split into the symmetric and the skew-symmetric parts $\mathbf{H}(\mathbf{x}) = \mathbf{E}(\mathbf{x}) + \boldsymbol{\Omega}(\mathbf{x})$, i.e. the first order-strain $\mathbf{E}(\mathbf{x}) = sym \nabla \mathbf{U}(\mathbf{x})$ and the main rotation $\boldsymbol{\Omega}(\mathbf{x}) = skw \nabla \mathbf{U}(\mathbf{x})$, respectively. The stress is described by the first-order stress $\boldsymbol{\Sigma}(\mathbf{x}) = \Sigma_{ij} \mathbf{e}_i \otimes \mathbf{e}_j$, $(\Sigma_{ij} = \Sigma_{ji})$ by the second-order stress tensor $\boldsymbol{\mu}(\mathbf{x}) = \mu_{ijk} \mathbf{e}_i \otimes \mathbf{e}_j \otimes \mathbf{e}_k$, $(\mu_{ijk} = \mu_{ikj})$ and by the non-symmetric real stress tensor $\mathbf{T}(\mathbf{x}) = \boldsymbol{\Sigma}(\mathbf{x}) - \nabla \bullet \boldsymbol{\mu}(\mathbf{x})$.

## 3 Multi-scale kinematics and asymptotic solution of the heterogeneous problem

According to [1], the local displacement field $\mathbf{u}(\mathbf{x})$ is represented through an asymptotic expansion in terms of parameter $\varepsilon$ of the micro-displacement, whose representation in components is

$$u_i\left(\mathbf{x}, \boldsymbol{\xi} = \frac{\mathbf{x}}{\varepsilon}\right) = \left( u_i^*(\mathbf{x}) + \sum_{l=1}^{+\infty} \varepsilon^l \sum_{|q|=l} N_{ijq}^l(\boldsymbol{\xi}) \frac{\partial^{|q|}}{\partial \mathbf{x}^q} u_j^*(\mathbf{x}) \right)\Bigg|_{\boldsymbol{\xi}=\frac{\mathbf{x}}{\varepsilon}} =$$
$$= \left( u_i^*(\mathbf{x}) + \varepsilon N_{ijq_1}^1(\boldsymbol{\xi}) \frac{\partial u_j^*(\mathbf{x})}{\partial x_{q_1}} + \varepsilon^2 N_{ijq_1q_2}^2(\boldsymbol{\xi}) \frac{\partial^2 u_j^*(\mathbf{x})}{\partial x_{q_1} \partial x_{q_2}} + \ldots \right)\Bigg|_{\boldsymbol{\xi}=\frac{\mathbf{x}}{\varepsilon}}. \quad (2)$$

In (2) $q$ is a multi-index, $u_i^*(\mathbf{x})$ is a $\mathscr{L}$-periodic function that only depends on the macroscopic coordinate $\mathbf{x}$ and the functions $N_{ijq}^l(\boldsymbol{\xi})$ are $\mathcal{Q}$-periodic perturbation functions having vanishing mean value on the unit cell $\mathcal{Q}$, i.e. $\langle N_{ijq}^l(\boldsymbol{\xi}) \rangle = \frac{1}{\delta} \int_{\mathcal{Q}} N_{ijq}^l(\boldsymbol{\xi}) d\boldsymbol{\xi} = 0$ (normalization condition).

The macro-displacement field is defined as the average of the micro-displacement on the unit cell $\mathcal{Q}$

$$U_i(\mathbf{x}) \doteq \left\langle u_i\left(\mathbf{x}, \frac{\mathbf{x}}{\varepsilon} + \boldsymbol{\zeta}\right) \right\rangle = \frac{1}{\delta} \int_{\mathcal{Q}} \left( u_i^*(\mathbf{x}) + \sum_{l=1}^{+\infty} \varepsilon^l \sum_{|p|=l} N_{ijq}^l(\boldsymbol{\xi}+\boldsymbol{\zeta}) \frac{\partial^{|q|}}{\partial \mathbf{x}^q} u_j^*(\mathbf{x}) \right)\Bigg|_{\boldsymbol{\xi}=\frac{\mathbf{x}}{\varepsilon}} d\boldsymbol{\zeta}, \quad (3)$$

where $\boldsymbol{\zeta} \in \mathcal{Q}$ and the translation vector $\varepsilon \boldsymbol{\zeta} \in \mathcal{A}$ measure all the possible translations of the heterogeneous medium compared to the $\mathscr{L}$-periodic body force $f_t(\mathbf{x})$. This averaging operation is seen as an ensemble average that is based on the argument that the precise phase of the microstructure with respect to the body force is generally unknown and a family of translated microstructures should therefore be considered [9]. Noting that for any $\mathcal{Q}$-periodic function $g(\boldsymbol{\xi}+\boldsymbol{\zeta})$ it results $\langle g(\boldsymbol{\xi}+\boldsymbol{\zeta}) \rangle = \frac{1}{\delta} \int_{\mathcal{Q}} g(\boldsymbol{\xi}+\boldsymbol{\zeta}) d\boldsymbol{\xi} = \frac{1}{\delta} \int_{\mathcal{Q}} g(\boldsymbol{\xi}+\boldsymbol{\zeta}) d\boldsymbol{\zeta}$ and taking into account the normalization condition imposed on the perturbation functions $N_{ijq}^l(\boldsymbol{\xi})$, via the equation (3) one obtains $U_i(\mathbf{x}) = u_i^*(\mathbf{x})$. According to equation (3) and by virtue of the properties of $N_{ijq}^l(\boldsymbol{\xi})$, the higher-order strain components $\kappa_{ijp_1\ldots p_m}(\mathbf{x})$ are

obtained in the form

$$\kappa_{iq_1q_2...q_m}(\mathbf{x}) = \frac{\partial^m U_i}{\partial x_{q_1} \partial x_{q_2} ... \partial x_{q_m}} = \left\langle \frac{\partial^m}{\partial x_{q_1} \partial x_{q_2} ... \partial x_{q_m}} u_i\left(\mathbf{x}, \frac{\mathbf{x}}{\varepsilon} + \zeta\right) \right\rangle. \qquad (4)$$

Notably, for $m = 1, 2$ the displacement gradient and the second-order strain take the following form $H_{iq_1}(\mathbf{x}) = \left\langle \frac{\partial}{\partial x_{q_1}} u_i\left(\mathbf{x}, \frac{\mathbf{x}}{\varepsilon} + \zeta\right) \right\rangle = \frac{1}{\delta} \int_Q \frac{\partial}{\partial x_{q_1}} u_i(\mathbf{x}, \xi) d\xi$,

$\kappa_{iq_1q_2}(\mathbf{x}) = \left\langle \frac{\partial^2}{\partial x_{q_1} \partial x_{q_2}} u_i\left(\mathbf{x}, \frac{\mathbf{x}}{\varepsilon} + \zeta\right) \right\rangle = \frac{1}{\delta} \int_Q \frac{\partial^2}{\partial x_{q_1} \partial x_{q_2}} u_i(\mathbf{x}, \xi) d\xi$, respectively.

Substituting the micro-displacement field (2) in equation (1), being $U_i(\mathbf{x}) = u_i^*(\mathbf{x})$ and $\frac{\partial}{\partial x_j} u_i\left(\mathbf{x}, \xi = \frac{\mathbf{x}}{\varepsilon}\right) = \left(\frac{\partial u_i}{\partial x_j} + \frac{1}{\varepsilon} \frac{\partial u_i}{\partial \xi_j}\right)\Big|_{\xi = \frac{\mathbf{x}}{\varepsilon}} = \left(\frac{\partial u_i}{\partial x_j} + \frac{u_{i,j}}{\varepsilon}\right)\Big|_{\xi = \frac{\mathbf{x}}{\varepsilon}}$, the equilibrium equation at the micro-scale may be rewritten as

$$\begin{aligned}&\left(\frac{1}{\varepsilon}\left[\left(C_{tlis}^m N_{ipq_1,s}^1\right)_{,l} + C_{tlpq_1,l}^m\right] H_{pq_1}(\mathbf{x}) + \right.\\&+ \left\{\left(C_{tlis}^m N_{ipq_1q_2,s}^2\right)_{,l} + \frac{1}{2}\left[\left(C_{tliq_2}^m N_{ipq_1}^1\right)_{,l} + C_{tq_2is}^m N_{ipq_1,s}^1 + C_{tq_2pq_1}^m + \left(C_{tliq_1}^m N_{ipq_2}^1\right)_{,l} + \right.\right.\\&\left.\left.+ C_{tq_1is}^m N_{ipq_2,s}^1 + C_{tq_1pq_2}^m\right]\right\} \kappa_{pq_1q_2}(\mathbf{x}) + .........\Big)\Big|_{\xi = \frac{\mathbf{x}}{\varepsilon}} = -f_t(\mathbf{x}), \qquad t = 1, 2.\end{aligned} \qquad (5)$$

In order to obtain an equilibrium equation in term of the macroscopic state variables alone (that depend on the slow coordinate), namely a PDE with constant coefficients, the unknown functions $N_{ipq}^1$, $N_{ipqr}^2$ and $N_{ipq_1...q_m}^m$ have to satisfy the following non-homogeneous equations (*cell problems*)

$$\begin{aligned}&\left(C_{tlis}^m N_{ipq_1,s}^1\right)_{,l} = -C_{tlpq_1,l}^m + \tilde{h}_{tpq_1} = -f_{tpq_1}^1 \\&\left(C_{tlis}^m N_{ipq_1q_2,s}^2\right)_{,l} = -\frac{1}{2}\left[\left(C_{tliq_2}^m N_{ipq_1}^1\right)_{,l} + C_{tq_2is}^m N_{ipq_1,s}^1 + C_{tq_2pq_1}^m + \left(C_{tliq_1}^m N_{ipq_2}^1\right)_{,l} + C_{tq_1is}^m N_{ipq_2,s}^1 + \right.\\&\qquad\qquad\qquad\qquad\left. + C_{tq_1pq_2}^m\right] + \tilde{h}_{tq_2pq_1} = -f_{tq_2pq_1}^2 \\&\qquad\qquad\qquad\qquad\vdots \\&\left(C_{tlis}^m N_{ipq_1...q_m,s}^m\right)_{,l} = -\frac{1}{m!} \sum_{\wp(q)}\left[\left(C_{tliq_m}^m N_{ipq_1...q_{m-1}}^{m-1}\right)_{,l} + C_{tq_mis}^m N_{ipq_1...q_{m-1},s}^{m-1} + C_{tq_miq_{m-1}}^m N_{ipq_1...q_{m-2}}^{m-2}\right] + \\&\qquad\qquad\qquad\qquad + \tilde{h}_{tq_mq_{m-1}pq_1...q_{m-2}} = -f_{tq_mq_{m-1}pq_1...q_{m-2}}^m\end{aligned} \qquad (6)$$

where the symbol $\wp(q)$ denotes all the possible permutations of the multi-index $q$ and the constants $\tilde{h}_{tpq_1}$, $\tilde{h}_{tq_2pq_1}$ and $\tilde{h}_{tq_mq_{m-1}pq_1\ldots q_{m-2}}$ are defined as follows

$$\tilde{h}_{tpq_1} = 0, \quad \tilde{h}_{tq_2pq_1} = \frac{1}{2}\left\langle C_{tq_2pq_1}^m + C_{tq_2is}^m N_{ipq_1,s}^1 + C_{tq_1pq_2}^m + C_{tq_1is}^m N_{ipq_2,s}^1 \right\rangle,$$

$$\tilde{h}_{tq_mq_{m-1}pq_1\ldots q_{m-2}} = \frac{1}{m!}\sum_{\wp(q)}\left\langle C_{tq_mis}^m N_{ipq_1\ldots q_{m-1},s}^{m-1} + C_{tq_miq_{m-1}}^m N_{ipq_1\ldots q_{m-2}}^{m-2} \right\rangle. \tag{7}$$

These constants are determined, in accordance with the asymptotic approach [1], so as to make the auxiliary body forces $f_{tpq}^1$, $f_{trpq}^2$ and $f_{tq_mq_{m-1}pq_1\ldots q_{m-2}}^m$ in equations (6) with vanishing mean value over the unit cell $Q$, i.e. $\langle f_{tpq}^1 \rangle = 0$, $\langle f_{tpqr}^2 \rangle = 0$ and $\langle f_{tq_mq_{m-1}pq_1\ldots q_{m-2}}^m \rangle = 0$. This assumption guarantees the $Q$-periodicity of the perturbation functions $N_{ipq}^j$ obtained as solution of problem (6) (see [1]) and implies the continuity of the micro-displacement fields and the anti-periodicity of the traction at the interface of adjacent cells. In accordance with (6) and (7), equation (5) is transformed into the average equation of infinite order introduced in [1], which takes the form

$$\tilde{h}_{kp_1jq_1}\kappa_{jq_1p_1} + \sum_{n=0}^{+\infty}\varepsilon^{n+1}\sum_{\substack{|q|=n+1\\|p|=2}}\tilde{h}_{kpjq}\kappa_{jqp} = -f_k. \tag{8}$$

It is well known that the truncation of equation (8) at a suitable order with the aim to obtain higher order field equations may generally lead to problems in which the symmetry of the higher-order elastic moduli $\tilde{h}_{tq_mq_{m-1}pq_1\ldots q_{m-2}}$ is not generally guaranteed and with possible loss of ellipticity of the differential problem, as observed in [9]. To circumvent these problems the asymptotic-variational homogenization techniques [9] may be appropriate. Alternatively, it is possible to directly determine the overall elastic moduli of the second-order homogenized continuum through a generalised macro-homogeneity condition based on an asymptotic expansion of the strain energy of the periodic domain $\mathscr{L}=[0,L]\times[0,\delta L]$ as described in the following Section. The here proposed technique preserves the simplicity of the computational approaches (based on the Hill-Mandel lemma i.e. an equality between the strain energy density to the two scales) and allows to determine the overall elastic moduli in terms of the perturbation functions $N_{ipq}^1$, $N_{ipqr}^2$, by including in the second-order moduli the contributions associated with the third-order strain tensor $\kappa_{pqrj}$.

The overall higher-order moduli are an approximation to those determined through the variational-asymptotic approach [9] but their determination is computationally less burdensome as discussed in Section 4. The generalized macro-homogeneity condition, here proposed, allows to analyze the correlations between the asymptotic and computational models here proposed (Section 4 and 5) and other approaches obtained in [8] and [18,22] through a comparison of the overall elastic moduli derived by several approaches.

## 4 Asymptotic expansion of the strain energy and second-order homogenization

An approximation $\mathbf{u}^{II}(\mathbf{x})$ of the micro-displacement $\mathbf{u}(\mathbf{x})$ given in (2) is assumed by retaining only the first two terms of the series, i.e.

$$u_i^{II}(\mathbf{x}) = u_i^{II}\left(\mathbf{x}, \boldsymbol{\xi} = \frac{\mathbf{x}}{\varepsilon}\right) = \left(U_i(\mathbf{x}) + \varepsilon N^1_{ipq_1}(\boldsymbol{\xi}) H_{pq_1}(\mathbf{x}) + \varepsilon^2 N^2_{ipq_1q_2}(\boldsymbol{\xi}) \kappa_{pq_1q_2}(\mathbf{x})\right)\bigg|_{\boldsymbol{\xi}=\frac{\mathbf{x}}{\varepsilon}}, \quad (9)$$

so that one obtains the components of the micro-displacement gradient

$$\frac{\partial u_i^{II}(\mathbf{x})}{\partial x_j} = \tilde{B}^H_{ijpq}\left(\boldsymbol{\xi} = \frac{\mathbf{x}}{\varepsilon}\right) H_{pq} + \varepsilon \tilde{B}^\kappa_{ijpqr}\left(\boldsymbol{\xi} = \frac{\mathbf{x}}{\varepsilon}\right) \kappa_{pqr} + \varepsilon^2 \tilde{A}^\kappa_{ipqr}\left(\boldsymbol{\xi} = \frac{\mathbf{x}}{\varepsilon}\right) \kappa_{pqrj}, \quad (10)$$

which also depends on the third-order strain $\kappa_{pqrj}$ components and where the localization tensors take the form

$$\tilde{B}^H_{ijpq}\left(\boldsymbol{\xi} = \frac{\mathbf{x}}{\varepsilon}\right) = \delta_{ip}\delta_{jq} + N^1_{ipq,j}, \quad \tilde{B}^\kappa_{ijpqr}\left(\boldsymbol{\xi} = \frac{\mathbf{x}}{\varepsilon}\right) = \frac{1}{2}\left(N^1_{ipq}\delta_{jr} + N^1_{ipr}\delta_{jq}\right) + N^2_{ipqr,j},$$
$$\tilde{A}^\kappa_{ipqr}\left(\boldsymbol{\xi} = \frac{\mathbf{x}}{\varepsilon}\right) = N^2_{ipqr}. \quad (11)$$

The translation $\varepsilon\boldsymbol{\zeta} \in \mathcal{A}$ of the periodic pattern (shown in Fig. 1.a) with respect to the $\mathcal{L}$-periodic body force implies, in general, a variation of the strain energy of the periodic domain $\mathcal{L}$. The strain energy is therefore influenced by the choice of periodic cell $\mathcal{A}$ which makes up the domain $\mathcal{L}$. These circumstances can be verified using the Fourier series representation of the strain energy density (represented as a function with slow and fast variables separated) and integrating term by term on $\mathcal{L}$. According to the previous observations, the elasticity tensor of the heterogeneous medium may be represented in the form $\mathbb{C}^{m,\zeta}\left(\boldsymbol{\xi} = \frac{\mathbf{x}}{\varepsilon}\right) = \mathbb{C}^m(\boldsymbol{\xi}+\boldsymbol{\zeta})$. Moreover, from equations (6), (7) and (11) it follows that both the perturbation functions $N^1_{ipq}$, $N^2_{ipqr}$, $N^m_{ipq_1\ldots q_m}$ and the components of the localization tensors $\tilde{B}_{ijpq}^{H}$, $\tilde{B}_{ijpqr}^{\kappa}$, $\tilde{A}_{ipqr}^{\kappa}$ depend on the parameter $\zeta$. Therefore, the strain energy density may be written

$$\phi_m^{\zeta}\left(\mathbf{x},\frac{\mathbf{x}}{\varepsilon}\right) = \phi_m\left(\mathbf{x},\frac{\mathbf{x}}{\varepsilon}+\zeta\right) = \left(\frac{1}{2}\frac{\partial u_i^{II}}{\partial x_j}C_{ijpq}^m\frac{\partial u_p^{II}}{\partial x_q}\right) = \\ = \left[\frac{1}{2}C_{ijkl}^m\left(\tilde{B}_{ijpq}^{H}H_{pq} + \varepsilon\tilde{B}_{ijpqr}^{\kappa}\kappa_{pqr} + \varepsilon^2\tilde{A}_{ipqr}^{\kappa}\kappa_{pqrj}\right)\left(\tilde{B}_{klp_1q_1}^{H}H_{p_1q_1} + \varepsilon\tilde{B}_{klp_1q_1r_1}^{\kappa}\kappa_{p_1q_1r_1} + \varepsilon^2\tilde{A}_{kp_1q_1r_1}^{\kappa}\kappa_{p_1q_1r_1l}\right)\right], \quad (12)$$

and the resulting strain energy $\mathcal{E}_m^{\zeta} = \mathcal{E}_m(\zeta)$ referred of domain $\mathcal{L}$ is defined as

$$\mathcal{E}_m^{\zeta} = \mathcal{E}_m(\zeta) = \int_{\mathcal{L}} \phi_m^{\zeta}\left(\mathbf{x},\frac{\mathbf{x}}{\varepsilon}\right)d\mathbf{x} = \int_{\mathcal{L}} \phi_m\left(\mathbf{x},\frac{\mathbf{x}}{\varepsilon}+\zeta\right)d\mathbf{x}, \quad (13)$$

To make the strain energy $\mathcal{E}_m^{\zeta}$ and the elastic moduli of the homogeneous continuum independent on the parameter $\zeta$, the strain energy is averaged with respect to all the possible realizations and the *mean strain energy* is introduced

$$\mathcal{E}_m \doteq \langle\mathcal{E}_m^{\zeta}\rangle = \frac{1}{\delta}\int_Q \mathcal{E}_m^{\zeta}d\zeta = \int_{\mathcal{L}}\left\langle\phi_m^{\zeta}\left(\mathbf{x},\frac{\mathbf{x}}{\varepsilon}\right)\right\rangle d\mathbf{x}, \quad (14)$$

where $\langle\bullet\rangle = \frac{1}{\delta}\int_Q \bullet d\xi = \frac{1}{\delta}\int_Q \bullet d\zeta$. From the $Q$-periodicity of the functions $C_{ijkl}^m(\xi+\zeta)$, $N_{ipq}^1(\xi+\zeta)$ and $N_{ipqr}^2(\xi+\zeta)$ and the localization tensors $\tilde{B}_{ijpq}^{H}$, $\tilde{B}_{ijpqr}^{\kappa}$ and $\tilde{A}_{ipqr}^{\kappa}$ defined in (11), the *mean strain energy* takes the form

$$\mathcal{E}_m = \frac{1}{2}\left\langle C_{ijkl}^m\tilde{B}_{ijpq}^{H}\tilde{B}_{klp_1q_1}^{H}\right\rangle\int_{\mathcal{L}}H_{pq}H_{p_1q_1}d\mathbf{x} + \varepsilon\left\langle C_{ijkl}^m\tilde{B}_{ijpq}^{H}\tilde{B}_{klp_1q_1r_1}^{\kappa}\right\rangle\int_{\mathcal{L}}H_{pq}\kappa_{p_1q_1r_1}d\mathbf{x} + \\ +\frac{\varepsilon^2}{2}\left[\left\langle C_{ijkl}^m\tilde{B}_{ijpqr}^{\kappa}\tilde{B}_{klp_1q_1r_1}^{\kappa}\right\rangle\int_{\mathcal{L}}\kappa_{pqr}\kappa_{p_1q_1r_1}d\mathbf{x} + \left\langle C_{stij}^m\tilde{A}_{ipqr}^{\kappa}\tilde{B}_{stkl}^{H} + C_{sjit}^m\tilde{B}_{itkl}^{H}\tilde{A}_{spqr}^{\kappa}\right\rangle\int_{\mathcal{L}}\kappa_{pqrj}H_{kl}d\mathbf{x}\right] + \quad (15) \\ +\frac{\varepsilon^3}{2}\left\langle C_{ijkl}^m\tilde{A}_{ipqr}^{\kappa}\tilde{B}_{klp_1q_1r_1}^{\kappa} + C_{ilkj}^m\tilde{B}_{ilp_1q_1r_1}^{\kappa}\tilde{A}_{kpqr}^{\kappa}\right\rangle\int_{\mathcal{L}}\kappa_{pqrj}\kappa_{p_1q_1r_1}d\mathbf{x} + \frac{\varepsilon^4}{2}\left\langle C_{ijkl}^m\tilde{A}_{ipqr}^{\kappa}\tilde{A}_{kp_1q_1r_1}^{\kappa}\right\rangle\int_{\mathcal{L}}\kappa_{pqrj}\kappa_{p_1q_1r_1l}d\mathbf{x}.$$

By neglecting the third and fourth order terms of $\varepsilon$, the divergence theorem is applied and the fourth term in equation (15) is properly symmetrized with respect to the repeated indices $q,r,j$, so obtaining

$$\mathcal{E}_m = \frac{1}{2}\left\langle C^m_{ijkl}\tilde{B}^H_{ijpq}\tilde{B}^H_{klp_1q_1}\right\rangle \int_{\mathcal{L}} H_{pq}H_{p_1q_1}d\mathbf{x} + \varepsilon\left\langle C^m_{ijkl}\tilde{B}^H_{ijpq}\tilde{B}^\kappa_{klp_1q_1r_1}\right\rangle \int_{\mathcal{L}} H_{pq}\kappa_{p_1q_1r_1}d\mathbf{x} +$$
$$+\frac{\varepsilon^2}{2}\left\{\left\langle C^m_{ijkl}\tilde{B}^\kappa_{ijpqr}\tilde{B}^\kappa_{klp_1q_1r_1}\right\rangle \int_{\mathcal{L}} \kappa_{pqr}\kappa_{p_1q_1r_1}d\mathbf{x} + \frac{1}{3}\left\langle C^m_{stij}\tilde{A}^\kappa_{ipqr}\tilde{B}^H_{stkl} + C^m_{sjit}\tilde{B}^H_{itkl}\tilde{A}^\kappa_{spqr}\right\rangle \left[\int_{\partial\mathcal{L}}\left(\kappa_{pqr}H_{kl}n_j + \right.\right.\right. \quad (16)$$
$$\left.\left.\left.+\kappa_{pqj}H_{kl}n_r + \kappa_{prj}H_{kl}n_q\right)ds - \int_{\mathcal{L}}\left(\kappa_{pqr}\kappa_{klj} + \kappa_{pqj}\kappa_{klr} + \kappa_{prj}\kappa_{klq}\right)d\mathbf{x}\right]\right\},$$

where the line integral along the boundary $\partial\mathcal{L}$ is vanishing as a consequence of the $\mathcal{L}$-periodicity of both the macro-displacement and of the macro-strain tensor. It follows that (in the light of equation (33) in Appendix A) the *mean strain energy* may be written as

$$\mathcal{E}_m = \frac{1}{2}\left\langle C^m_{ijkl}B^H_{ijpq}B^H_{klp_1q_1}\right\rangle \int_{\mathcal{L}} H_{pq}H_{p_1q_1}d\mathbf{x} + \varepsilon\left\langle C^m_{ijkl}B^H_{ijpq}B^\kappa_{klp_1q_1r_1}\right\rangle \int_{\mathcal{L}} H_{pq}\kappa_{p_1q_1r_1}d\mathbf{x} + \\ +\frac{\varepsilon^2}{2}\left\langle C^m_{ijkl}B^\kappa_{ijpqr}B^\kappa_{klp_1q_1r_1}\right\rangle \int_{\mathcal{L}} \kappa_{pqr}\kappa_{p_1q_1r_1}d\mathbf{x} - \frac{\varepsilon^2}{24}\left\langle A^{H\_\kappa}_{p_1q_1r_1p_2q_2r_2}\right\rangle \int_{\mathcal{L}} \kappa_{p_1q_1r_1}\kappa_{p_2q_2r_2}d\mathbf{x}, \quad (17)$$

where the localization tensors $B^H_{ijpq}$, $B^\kappa_{ijpqr}$ and $A^{H\_\kappa}_{p_1q_1r_1p_2q_2r_2}$ are shown in Appendix A. Notably, in the tensor $A^{H\_\kappa}_{p_1q_1r_1p_2q_2r_2}$ are collected the contributions of the micro-displacement gradient associated to the macro-displacement gradient $H_{kl}$ and to the third-order strain tensor $\kappa_{pqrj}$, first and third term, respectively, of the equation (10).

In the second-order continuum the generalised macro-strain energy of the domain $\mathcal{L}$ is written as

$$\mathcal{E}_M = \int_{\mathcal{L}} \phi_M(\mathbf{x})d\mathbf{x} = \frac{1}{2}C_{ijkl}\int_{\mathcal{L}} H_{ij}H_{kl}d\mathbf{x} + Y_{ijklp}\int_{\mathcal{L}} H_{ij}\kappa_{klp}d\mathbf{x} + \frac{1}{2}S_{ijrklp}\int_{\mathcal{L}} \kappa_{ijr}\kappa_{klp}d\mathbf{x}, \quad (18)$$

$\phi_M$ being the strain energy density. From the generalised macro-homogeneity condition $\mathcal{E}_m = \mathcal{E}_M$, the overall elastic moduli of the homogenized continuum are obtained

$$C_{pqp_1q_1} = \left\langle C^m_{ijkl}B^H_{ijpq}B^H_{klp_1q_1}\right\rangle, \quad \frac{Y_{pqp_1q_1r_1}}{\varepsilon} = \left\langle C^m_{ijkl}B^H_{ijpq}B^\kappa_{klp_1q_1r_1}\right\rangle, \quad \frac{S_{pqrp_1q_1r_1}}{\varepsilon^2} = \left\langle C^m_{ijkl}B^\kappa_{klp_1q_1r_1}B^\kappa_{ijpqr}\right\rangle - \frac{\left\langle A^{H\_\kappa}_{pqrp_1q_1r_1}\right\rangle}{12}, (19)$$

which result to be not affected by the choice of the periodic cell $\mathcal{A}$. In fact, the $\zeta$-average operation denoted by $\langle\cdot\rangle$ is carried out with respect to all the possible realizations obtained by shift of the periodic pattern with respect to the $\mathcal{L}$-periodic body force. Moreover, the properties of the cell problems (6) that ensure the $Q$-periodicity of the perturbation functions $N^m_{ipq_1...q_m}$, allow to choose indifferently or the minimal cell or a cluster of minimal cells, what periodic cell $\mathcal{A}$ that constitutes the periodic domain $\mathcal{L}$. The overall elastic moduli are an approximation of those determined through the variational-asymptotic approach [9]. This approximation is due to the truncation of the micro-displacement field that is approximated by a second order asymptotic expansion (see equation (9)). On the other hand in the variational-asymptotic method, the micro-displacement field is represented through the asymptotic expansion (2) and the micro-displacement gradient is truncated to the third order. As a consequence the micro-displacement gradient also depends by the perturbation functions $N^3_{ipqrs}$. The determination of these sixteen additional functions is in general rather laborious and computationally burdensome because they are solution of non-homogeneous cell problems with body forces that depend by the functions $N^1_{ipq}$ and $N^2_{ipqr}$.

In the limit case of locally homogeneous material (microstructure disappearing), i.e. of vanishing mismatch of the elastic moduli of the single phases, the perturbation functions $N^m_{ipq_1...q_m}$ are vanishing and the localization tensors take the form

$$B^H_{ijpq}\left(\boldsymbol{\xi}=\frac{\mathbf{x}}{\varepsilon}\right)=\frac{1}{4}\left(\delta_{ip}\delta_{jq}+\delta_{jp}\delta_{iq}+\delta_{iq}\delta_{jp}+\delta_{jq}\delta_{ip}\right), \quad B^\kappa_{ijpqr}\left(\boldsymbol{\xi}=\frac{\mathbf{x}}{\varepsilon}\right)=A^{H\_\kappa}_{p_1q_1r_1p_2q_2r_2}\left(\boldsymbol{\xi}=\frac{\mathbf{x}}{\varepsilon}\right)=0. \quad (20)$$

In this case the elastic moduli from equation (19) turn out to be those of the classical homogenized continuum:

$$C_{pqp_1q_1}=\left\langle C^m_{ijkl}B^H_{ijpq}B^H_{klp_1q_1}\right\rangle=C^m_{pqp_1q_1}, \quad Y_{pqp_1q_1r_1}=0, \quad S_{pqrp_1q_1r_1}=0, \quad (21)$$

independently of the characteristic size $\varepsilon$ of the periodic cell.

A coupling between the macro-stress tensors $\Sigma_{pq}$, $\mu_{pqr}$ and the micro-stress tensor $\sigma_{ij}$ in the unit cell is deduced from an application of the generalised macro-homogeneity condition, referred to the domain $\mathcal{L}$. By equations (10) and (17) the mean strain energy at microscale may be written in the form

$$\begin{aligned}\mathcal{E}_m &= \frac{1}{2}\int_{\mathcal{L}}\left\langle C^m_{ijkl}B^H_{ijpq}\left[B^H_{klp_1q_1}H_{p_1q_1}+\varepsilon B^\kappa_{klp_1q_1r_1}\kappa_{p_1q_1r_1}+O(\varepsilon^2)\right]\right\rangle H_{pq}d\mathbf{x}+ \\ &+\frac{1}{2}\int_{\mathcal{L}}\left\langle \varepsilon C^m_{ijkl}B^\kappa_{klp_1q_1r_1}\left[B^H_{klpq}H_{pq}+\varepsilon B^\kappa_{klpqr}\kappa_{pqr}+O(\varepsilon^2)\right]\right\rangle \kappa_{p_1q_1r_1}d\mathbf{x}+O(\varepsilon^3)\end{aligned} \quad (22)$$

and by applying the generalised macro-homogeneity condition the macro-stress tensors $\Sigma_{pq}$ and $\mu_{pqr}$ are obtained, with the terms of order $O(\varepsilon^3)$ neglected, in terms of the micro-stress field, i.e.

$$\Sigma_{pq} = \langle B^H_{ijpq} \sigma^{II}_{ij} \rangle, \qquad \mu_{pqr} = \varepsilon \langle B^\kappa_{ijpqr} \sigma^{II}_{ij} \rangle, \qquad (23)$$

where $\sigma^{II}_{ij} = C^m_{ijkl} \left[ B^H_{klp_1q_1} H_{p_1q_1} + \varepsilon B^\kappa_{klp_1q_1r_1} \kappa_{p_1q_1r_1} \right] + O(\varepsilon^2)$. If one assumes the micro-stress $\sigma_{ij}$ approximated by the stress field $\sigma^{II}_{ij}$ resulting by the down-scaling, i.e. $\sigma_{ij} \approx \sigma^{II}_{ij}$, the macro-stress tensors given by equation (23), taking into account equation (34), are written as follows

$$\Sigma_{pq} = \langle \sigma_{pq} \rangle + \frac{\varepsilon}{2} \frac{\partial}{\partial x_j} \langle \sigma_{ij} (N^1_{ipq} + N^1_{iqp}) \rangle,$$

$$\mu_{pqr} = \frac{\varepsilon}{2} \langle \sigma_{ir} N^1_{ipq} + \sigma_{iq} N^1_{ipr} \rangle + \varepsilon^2 \frac{\partial}{\partial x_j} \langle \sigma_{ij} N^2_{ipqr} \rangle, \qquad (24)$$

and the real stress tensor components are

$$T_{pq} = \Sigma_{pq} - \frac{\partial \mu_{pqr}}{\partial x_r} = \langle \sigma_{pq} \rangle + \frac{\varepsilon}{2} \frac{\partial}{\partial x_j} \langle \sigma_{ij} N^1_{iqp} - \sigma_{iq} N^1_{ipj} \rangle - \varepsilon^2 \frac{\partial^2}{\partial x_j \partial x_r} \langle \sigma_{ij} N^2_{ipqr} \rangle. \qquad (25)$$

It is worth to note that by definitions (24) and (25) the macroscopic balance equation of the second-order continuum $\frac{\partial T_{pq}}{\partial x_q} + f_p = \frac{\partial}{\partial x_q} \left( \Sigma_{pq} - \frac{\partial \mu_{pqr}}{\partial x_r} \right) + f_p = 0$ is obtained by averaging of the microscopic balance equation $\frac{\partial \sigma_{pq}}{\partial x_q} + f_p = 0$ over the periodic cell. In fact, by truncating the real stress (25) to the second order and applying the divergence theorem, by virtue of the Q-periodicity of the stress $\sigma_{pq}$, one obtains $\frac{\partial T_{pq}}{\partial x_q} = \frac{\partial \langle \sigma_{pq} \rangle}{\partial x_q}$. This representation for the macro-stresses components (24) have an analogous form to those obtained in [8] for strain gradient continuum.

Finally, it is worth to note that the homogenization based on a reduced second-order continuum such as the Koiter equivalent continuum is possible according to the considered approach, as shown in Appendix B.

# 5 Computational homogenizations: polynomial approximation of the macro-displacement field and generalised periodic boundary conditions of the unit cell

To establish a correlation between the asymptotic approach and the computational homogenization methods (based on the *quadratic ansätze* [19]), an approximation of the macro-displacement field through a second-order Taylor polynomial expansion is considered. In this case the macro-displacement, the first- and second-order strain components may be written as

$$U_p(\mathbf{x}) = \bar{U}_p + \bar{H}_{pq} x_q + \frac{1}{2}\bar{\kappa}_{pqr} x_q x_r, \quad H_{pq}(\mathbf{x}) = \bar{H}_{pq} + \frac{1}{2}\left(\bar{\kappa}_{pqr} x_r + \bar{\kappa}_{prq} x_r\right), \quad \kappa_{pqr}(\mathbf{x}) = \bar{\kappa}_{pqr}, \quad (26)$$

$\bar{U}_p$, $\bar{H}_{pq}$ and $\bar{\kappa}_{pqr}$ being the macro-displacement and the first and second displacement gradient evaluated at point $\mathbf{x} = \mathbf{0}$, respectively. As a consequence, the micro-displacement field approximation takes the form

$$u_i^{II}(\mathbf{x}) = u_i^{II}\left(\mathbf{x}, \boldsymbol{\xi} = \frac{\mathbf{x}}{\varepsilon}\right) = \bar{U}_p + \bar{H}_{pq} x_q + \frac{1}{2}\bar{\kappa}_{pqr} x_q x_r + \left(\frac{\varepsilon}{2}\left[N_{ipq}^1(\boldsymbol{\xi}) x_r + N_{ipr}^1(\boldsymbol{\xi}) x_q\right]\bar{\kappa}_{pqr} + \right. \\ \left. + \varepsilon N_{ipq}^1(\boldsymbol{\xi})\bar{H}_{pq} + \varepsilon^2 N_{ipqr}^2(\boldsymbol{\xi})\bar{\kappa}_{pqr}\right)\Big|_{\boldsymbol{\xi} = \frac{\mathbf{x}}{\varepsilon}}, \quad (27)$$

that justifies from a different point of view the micro-displacement field assumed in a heuristic way in [18]. Notably, the terms linearly depending on parameter $\varepsilon$ are proportional to both the displacement gradient $\bar{H}_{pq}$ and the second gradient $\bar{\kappa}_{pqr}$. Moreover, the micro-displacement gradient

$$\frac{\partial u_i^{II}(\mathbf{x})}{\partial x_j} = \tilde{B}_{ijpq}^H\left(\boldsymbol{\xi} = \frac{\mathbf{x}}{\varepsilon}\right) H_{pq} + \varepsilon \tilde{B}_{ijpqr}^\kappa\left(\boldsymbol{\xi} = \frac{\mathbf{x}}{\varepsilon}\right) \kappa_{pqr}, \quad (28)$$

differs from the one given by expansion (10) only with regard to the term associated with the macro-strain $\kappa_{pqrj}$, that is assumed vanishing in the truncated expansions (26).

From equations (15), (17) and (27) the mean strain energy $\mathscr{E}_m$ is written as

$$\mathscr{E}_m = \frac{1}{2}\left\langle C_{ijkl}^m B_{ijpq}^H B_{klp_1q_1}^H \right\rangle \int_\mathscr{L} H_{pq} H_{p_1q_1} d\mathbf{x} + \varepsilon \left\langle C_{ijkl}^m B_{ijpq}^H B_{klp_1q_1r_1}^\kappa \right\rangle \int_\mathscr{L} H_{pq} \kappa_{p_1q_1r_1} d\mathbf{x} + \\ + \frac{\varepsilon^2}{2}\left\langle C_{ijkl}^m B_{ijpqr}^\kappa B_{klp_1q_1r_1}^\kappa \right\rangle \int_\mathscr{L} \kappa_{pqr} \kappa_{p_1q_1r_1} d\mathbf{x}, \quad (29)$$

where $B_{ijpq}^{H}$ e $B_{ijpqr}^{\kappa}$ are the localization tensors, suitably symmetrised with respect to the saturated indices shown in (34). From the generalised macro-homogeneity condition $\mathcal{E}_m = \mathcal{E}_M$, being the macro-stain energy $\mathcal{E}_M$ defined in (18), the following approximations of the elastic moduli of the second-order continuum are obtained

$$C_{pqp_1q_1} = \left\langle C_{ijkl}^m B_{ijpq}^H B_{klp_1q_1}^H \right\rangle, \quad \frac{Y_{pqp_1q_1r_1}}{\varepsilon} = \left\langle C_{ijkl}^m B_{ijpq}^H B_{klp_1q_1r_1}^\kappa \right\rangle, \quad \frac{S_{pqrp_1q_1r_1}}{\varepsilon^2} = \left\langle C_{ijkl}^m B_{klp_1q_1r_1}^\kappa B_{ijpqr}^\kappa \right\rangle, \quad (30)$$

which differ from the corresponding ones given by equations (19) in the second-order moduli for the term associated with the localization tensor $A_{p_1q_1r_1p_2q_2r_2}^{H-\kappa}$. This difference is induced by the quadratic ansätze which is the basis of the computational approaches in which the macro-displacement is approximated through a second Taylor polynomial expansion (see equation (26)) in which the macro-strain $\kappa_{pqrj}$ are assumed zero. In the limit case of locally homogeneous material, i.e. of vanishing mismatch of the elastic moduli of the single phases the higher-order elastic moduli $Y_{pqp_1q_1r_1}$ and $S_{pqrp_1q_1r_1}$ are vanishing. The overall elastic moduli in the equation (30) have the same form of those obtained through different procedures in [22] and in [8] in the case of strain gradient continuum. In the latter, the overall elastic moduli are obtained by applying a generalization of Hill-Mandel lemma between the $\xi$-average of the micro-strain energy density (i.e. calculated as the average of the density $\phi_m(\mathbf{x}, \xi)$ over the area of the unit cell $Q$ with respect to the microscopic variable $\xi$) and the macro-strain energy density. The overall elastic moduli thus determined do not depend on the localization tensor $A_{p_1q_1r_1p_2q_2r_2}^{H-\kappa}$. In fact, such dependence is obtained by applying the generalized macro-homogeneity condition in which the *mean strain energy* $\mathcal{E}_m$ on the periodic domain $\mathcal{L}$ is set equal to the macro-strain energy $\mathcal{E}_M$ (see Section 4). The here proposed *mean strain energy* is defined via $\zeta$- and $\mathbf{x}$-average of the micro-strain energy density $\phi_m^\zeta$, expressed in (12), over the area of the unit cell $Q$ with respect to the variable $\zeta$ and over the area of the cluster $\mathcal{L}$ with respect to the variable $\mathbf{x}$. The $\mathbf{x}$-average operation allows the application of the divergence theorem (see equations (15) and (16)) by which is possible to include in the second order overall elastic moduli the terms associated to the third-order strain tensor. These contributions are collected in the localization tensor $A_{p_1q_1r_1p_2q_2r_2}^{H-\kappa}$ in equation (34) (see

Appendix A). Conversely, this theorem can not be directly applied to the $\xi$-average of the micro-strain energy density $\phi_m(\mathbf{x},\xi)$.

The corresponding homogenization of a multi-polar continuum [32] and of a Koiter continuum may be developed as shown in Appendices B and C, respectively.

A computational simplification of the homogenization technique previously described may be obtained by solving the equilibrium equations (1) in the periodic cell $\mathscr{A}$ in term of micro-displacement $\mathbf{u}^{II}(\mathbf{x})$ for vanishing body forces (as proposed in [18,22]). This simplification may be obtained through an approximation of the micro-displacement $\mathbf{u}^{II}(\mathbf{x})$ by preserving the formal structure of equation (27) and replacing the $Q$-periodic fluctuation functions $N_{ikl}^1(\xi)$ and $N_{iklp}^2(\xi)$ with fluctuation functions $\vartheta_{ikl}^1(\xi)$ and $\vartheta_{iklp}^2(\xi)$ which are the solutions of properly defined cell problems. The generalised $\mathscr{A}$-periodic boundary condition are

$$
\begin{aligned}
u_i^{II}\left(\varepsilon\xi_{b1}^+\right) - u_i^{II}\left(\varepsilon\xi_{b1}^-\right) &= U_i\left(\varepsilon\xi_{b1}^+\right) - U_i\left(\varepsilon\xi_{b1}^-\right) + \\
&+ \varepsilon\vartheta_{i11}^{1+}d_1\overline{\kappa}_{111} + \frac{\varepsilon}{2}\vartheta_{i12}^{1+}d_1\left(\overline{\kappa}_{112}+\overline{\kappa}_{121}\right) + \varepsilon\vartheta_{i21}^{1+}d_1\overline{\kappa}_{211} + \frac{\varepsilon}{2}\vartheta_{i22}^{1+}d_1\left(\overline{\kappa}_{221}+\overline{\kappa}_{212}\right), \quad \xi_{b2}\in C_1 \\
u_i^{II}\left(\varepsilon\xi_{b2}^+\right) - u_i^{II}\left(\varepsilon\xi_{b2}^-\right) &= U_i\left(\varepsilon\xi_{b2}^+\right) - U_i\left(\varepsilon\xi_{b2}^-\right) + \\
&+ \frac{\varepsilon}{2}\vartheta_{i11}^{1+}d_2\left(\overline{\kappa}_{112}+\overline{\kappa}_{121}\right) + \varepsilon\vartheta_{i12}^{1+}d_2\overline{\kappa}_{122} + \frac{\varepsilon}{2}\vartheta_{i21}^{1+}d_2\left(\overline{\kappa}_{212}+\overline{\kappa}_{221}\right) + \varepsilon\vartheta_{i22}^{1+}d_2\overline{\kappa}_{222}, \quad \xi_{b1}\in C_2
\end{aligned}
\tag{31}
$$

where $U_i = \overline{U}_i + \overline{H}_{iq}x_q + \frac{1}{2}\overline{\kappa}_{iqr}x_qx_r$, $\xi_{b1}^{\pm} = \pm\frac{1}{2}$, $\xi_{b2}^{\pm} = \pm\frac{\delta}{2}$, $\vartheta_{ipq}^{1+} = \vartheta_{ipq}^1\left(\xi_{bj}^+\right)$, $d_1 = \varepsilon$, $d_2 = \delta\varepsilon$ with $j = 1,2$ and $C_1$, $C_2$ being the vertical and horizontal boundaries of the unit cell $Q$, respectively.

This procedure follows from the approach proposed in [18,22] and is carried out in two successive steps: a first-order homogenization related to prescribed homogeneous macro-strain $\mathbf{H}$ and a second-order homogenization with prescribed homogeneous second-order macro-strain $\boldsymbol{\kappa}$. These strains are imposed on the cell through the generalised $\mathscr{A}$-periodic boundary condition (31). It is worth to note that the fluctuation functions $\vartheta_{ikl}^1(\xi)$ coincide with functions $N_{ikl}^1(\xi)$. On the contrary, the functions $\vartheta_{iklp}^2(\xi)$ differ in general from the functions $N_{iklp}^2(\xi)$. According to this approach, the micro-displacement field obtained by localization has the property of being continuous across the interface of adjacent periodic cells $\mathcal{A}$. However, the stress field results not to be continuous across the cell interface where the anti-periodicity of the traction is not guaranteed. Finally, the elastic moduli obtained by the above approximation present an analogous structure to those shown in (30), once in the localization tensors $B_{ijpq}^{H}$ and $B_{ijpqr}^{K}$ (equations (21.1) and (21.2)) the perturbations $N_{ikl}^{1}(\xi)$ and $N_{iklp}^{2}(\xi)$ are substituted by functions $\vartheta_{ikl}^{1}(\xi)$ and $\vartheta_{iklp}^{2}(\xi)$, respectively (see [22]).

## 6 Illustrative examples: homogenization of a three-phase composite

The homogenization techniques proposed in this paper have been applied to the analysis of an orthotropic three-phase composite and the results obtained are compared with those obtained by the homogenization methods proposed in [8] and [18,22]. This three-phase composite here analysed has been proposed in [18] in order to assess the reliability of results obtained through homogenization techniques because it presents characteristic lengths along the two orthotropic axes (see equation (32)) significantly different as will be shown in the following. The three-phase composite is characterized by a square periodic cell ($\delta = 1$) having size $\varepsilon$ and shown in Figure 2.b. The constituents are assumed isotropic, perfectly bonded and in plane stress condition, the ratios $\eta = E_3/E_2$, $\psi^{-1} = (E_3 + E_2)/2E_1$ between the Young's moduli of the phases are introduced and the Poisson ratios $\nu_1 = \nu_2 = \nu_3$ are assumed to be equal.

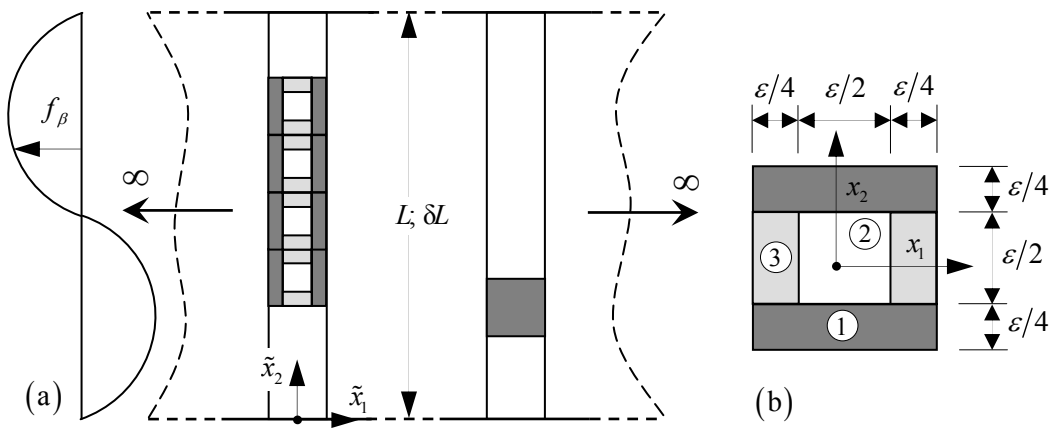

Figure 2: (a) Heterogeneous and homogenized models with $\mathcal{L}$-periodic body force; (b) Periodic cell and constituents: three-phase composite.

To evaluate the capabilities of the homogenization procedures, the two-dimensional orthotropic heterogeneous material is considered subjected to $\mathcal{L}$-periodic harmonic body forces $f_\beta$ along the orthotropy direction $\tilde{x}_\alpha$ (see figure 2.a), i.e. $f_\beta(\tilde{x}_\alpha) = \Xi_\beta\, e^{i\left(\frac{2\pi}{L_\alpha}\tilde{x}_\alpha\right)}$ with $\alpha,\beta = 1,2$, $L_1 = L_2 = L$, $\Xi_\beta \in \mathbb{R}$ and $i^2 = -1$. Two different models have been considered. The former is the heterogeneous one, where each phase is represented, and the latter is the homogenised second-order model with overall elastic moduli derived from the homogenization of the periodic cell through the approaches developed in Sections 4 and 5. Due to the periodicity of both the heterogeneous material and of the body forces considered, only a horizontal (or vertical) characteristic portion of length $L$ of the heterogeneous model is analysed (Figure 2.a). In order to assess the reliability of the second-order model, the macro-displacement at some meaningful unit cells in the homogenized model are compared to the corresponding ones in the heterogeneous model.

Once evaluated the perturbation functions $N^1_{ipq}$, $N^2_{ipqr}$ by a FE analysis of a periodic cell, the overall elastic moduli of the second-order equivalent continuum model are obtained both through the asymptotic approach (Section 4) and by the computational ones based on the quadratic assumption (26) and on the further simplification on the fluctuation functions [18] (Section 5).

The solution of the heterogeneous problem with $\mathcal{L}$-periodic harmonic body forces is computed via FE analysis with periodicity boundary condition on the displacement field, while the solution to the homogenized problem is given in Appendix D. The characteristic lengths $\lambda_{Sh}$ and $\lambda_{Ext}$ associated with the shear and the extensional strain along the directions $x_1$ and $x_2$ (see figure 2.b) depend on the overall elastic moduli according the following definitions:

$$\lambda_{Sh-1} = \lambda_1^2 = \sqrt{\frac{S_{211211}}{C_{1212}}}, \quad \lambda_{Sh-2} = \lambda_2^1 = \sqrt{\frac{S_{122122}}{C_{1212}}}, \quad \lambda_{Ext-1} = \lambda_1^1 = \sqrt{\frac{S_{111111}}{C_{1111}}}, \quad \lambda_{Ext-2} = \lambda_2^2 = \sqrt{\frac{S_{222222}}{C_{2222}}}, \quad (32)$$

and are represented in the dimensionless diagrams in Figure 3 as a function of the elastic mismatch ratio $\eta = E_3/E_2$ with $\psi = 2000/11$, $\nu_1 = \nu_2 = \nu_3 = 0.1$, for the cases of asymptotic approach (red line), the computational approach (blue line) and simplified computational model [18] (green line), respectively. These lengths are strongly influenced by the stiffness mismatches $\eta$, $\psi$ but weakly depend on the Poisson ratios. It is worth to note that the characteristic lengths evaluated by the computational approach which are (analogous to those obtained in [8] for strain gradient continuum) do not depend on the contribution associated with the localization tensor $A^{H-\kappa}_{p_1q_1r_1p_2q_2r_2}$ (see equation (30) compared to (19)) unlike those obtained by the asymptotic approach. The characteristic lengths evaluated through the computational homogenization methods are greater, for each value of ratio $\eta$, of those obtained through the more accurate asymptotic approach. Notably, those differences induced by the contributions associated with localization tensor $A^{H-\kappa}_{p_1q_1r_1p_2q_2r_2}$, may be significant in cases where the characteristic lengths (evaluated through the more accurate asymptotic approach) are small (i.e. $\lambda/\varepsilon < 0.2$) like obtained for $\lambda_{Ext}$ along $x_1$, $x_2$ (see Figure 3.b, 3.d) and for $\lambda_{Sh}$ along $x_2$ (see Figure 3.c) in terms of the stiffness mismatch $\eta$. Moreover, the contribution associated with $A^{H-\kappa}_{p_1q_1r_1p_2q_2r_2}$ in the second-order elastic moduli (equation (19)) may significantly affect the lowest (acoustic) branch of the dispersion curves (see [22,23]) for compression waves along $x_1$, $x_2$ and shear waves along $x_2$.

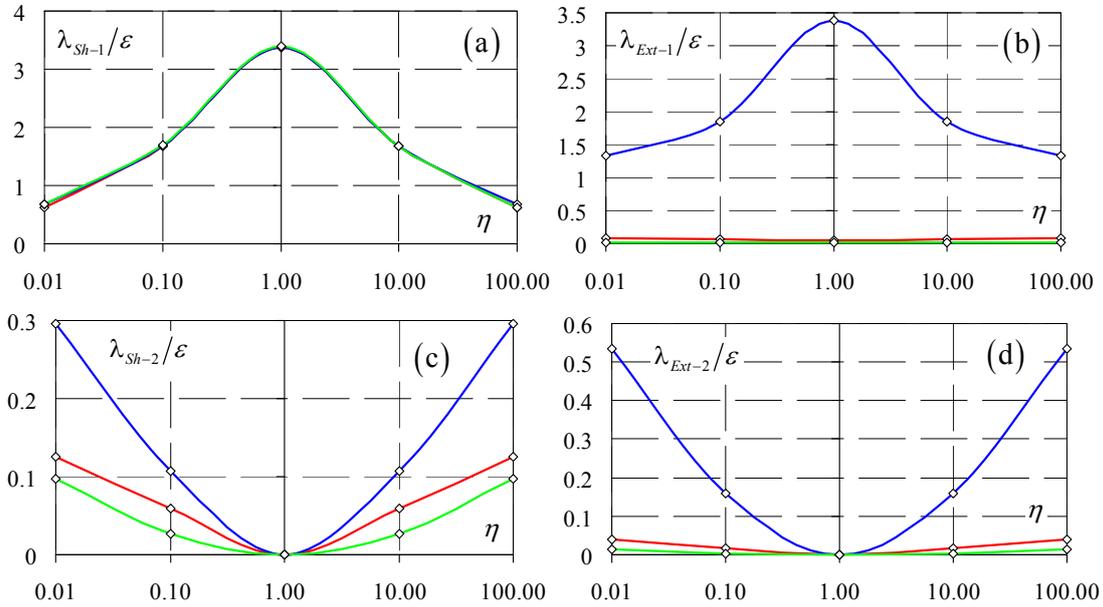

Figure 3: Characteristic shear and extensional lengths $\lambda_{Sh}$ and $\lambda_{Ext}$ versus stiffness mismatch $\eta$: red line asymptotic approach; blue line computational approach; green line simplified computational approach [18].

It is worth to note that in the simplified computational approach [18] (such as in other approaches in which the macro-displacement is described by a Taylor polynomial expansion [11-19]) the characteristic lengths may be influenced by the choice of the periodic cell of characteristic size $\varepsilon$. Nevertheless, this effect is minimal in the present example inasmuch $\lambda_{Sh}(\eta) \cong \lambda_{Sh}(1/\eta)$ $\left(\lambda_{Ext}(\eta) \cong \lambda_{Ext}(1/\eta)\right)$ (see Figure 3, green line).

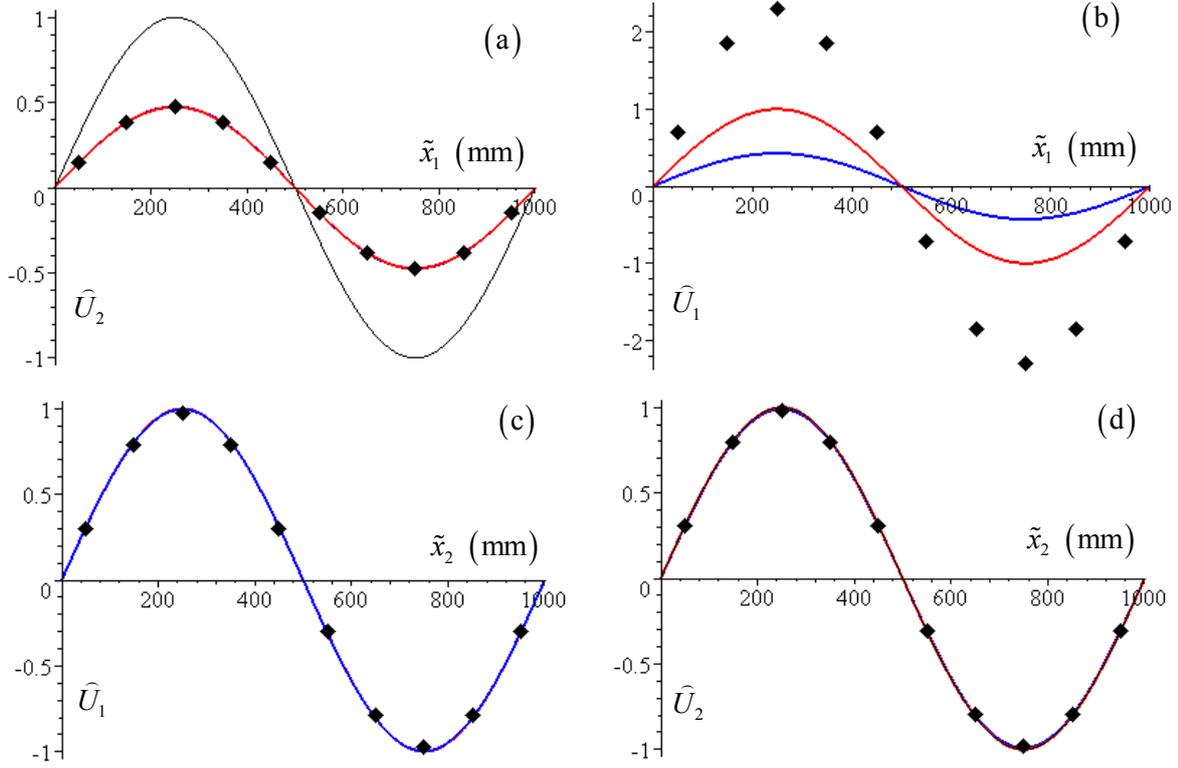

Figure 4: Displacement field due to harmonic body forces $f_\beta(\tilde{x}_\alpha)$: Shear problem $\alpha \neq \beta$, dimensionless macro-displacement: (a) $\widehat{U}_2(x_1)$, (c) $\widehat{U}_1(x_2)$. Extensional problem $\alpha = \beta$, macro-displacement: (b) $\widehat{U}_1(x_1)$, (d) $\widehat{U}_2(x_2)$. Heterogeneous model (diamonds) versus homogeneous second-order and first-order models (Second order model: red line asymptotic approach; blue line computational approach; green line simplified computational approach [18]. First order model black line).

The numerical results obtained from the heterogeneous models for $L/\varepsilon = 10$, $f_\beta(\tilde{x}_\alpha) = \Xi_\beta \, \Im\!\left[e^{i\left(\frac{2\pi}{L_\alpha}\tilde{x}_\alpha\right)}\right]$ (with $\Im[\bullet]$ imaginary part), $\Xi_\beta = 1$ N/mm$^3$, $\eta = E_3/E_2 = 10$, $\psi = 2000/11$ and $\nu_1 = \nu_2 = \nu_3 = 0.1$, are shown in the diagrams of Figure 4 as diamonds representing the dimensionless macro-displacement $\widehat{U}_\beta = \dfrac{4\pi^2 U_\beta C_{\beta\alpha\beta\alpha}}{\Xi_\beta L_\alpha^{\,2}}$ of the unit cells along the characteristic portion of length $L$ (with varying $\tilde{x}_\alpha$). The continuous red, blue and green lines in the diagrams represent the corresponding results from the homogenized second-order model with overall elastic moduli evaluated through the asymptotic, computational and simplified computational approaches, respectively. The black lines represent the results from the first-order model. From these diagrams a good agreement is observed between the results from the second-order models and those from the heterogeneous model for both the *shear problems* (Figure 4.a, 4.c) and the *extensional problem* along direction $x_2$ (Figure 4.d).

In the cases of shear and extensional problems along direction $x_2$, the non-local effects (i.e. the characteristic lengths $\lambda_{Sh-2}$ and $\lambda_{Ext-2}$) appear to be negligible in comparison with the case of shear problem along $x_1$. These results may be explained by the consideration that the three-phase composite analysed is similar to a layered material that has vanishing characteristic lengths along the normal direction of the layers [18]. Moreover, it can be seen how the diagrams obtained from the second-order model characterised through the asymptotic and computational models turn out to be almost coincident with that of the first-order homogeneous continuum (see Figures 4.c and 4.d). In fact, for small dimensionless characteristic lengths $\lambda_{Sh-2}/\varepsilon$ or $\lambda_{Ext-2}/\varepsilon$ (i.e. $\lambda/\varepsilon < 0.1$) and for $L/\varepsilon = 10$ (as assumed in the example with $\eta = E_3/E_2 = 10$) the macro-displacement obtained by non-local models is approximately equal to that obtained by the Cauchy continuum, i.e. $U_\beta \simeq U_\beta^C$ (see equation (45) in Appendix D and Figures 3.c, 3.d and 5.a). Therefore, although there are significant differences on the characteristic lengths obtained by the different approaches (see Figure 3), these differences are not instead detectable in the macro-displacement (or in the displacement gradient) for $L/\varepsilon = 10$ as shown in Figures 4.a, 4.c, 4.d. These differences are remarkable in cases in which the ratio $L/\varepsilon$ is sufficiently small, i.e. $L/\varepsilon = 2 \div 3$, or in the case of $\mathscr{L}$-periodic forces that have harmonic contents not negligible for the wave-numbers $2\pi n/L = 4\pi/L \div 6\pi/L$ (with $n = \pm L/\varepsilon$ and $n \in \mathbb{Z}$). In summary, the homogeneous second gradient continuum characterised by a) the asymptotic approach (Section 4), b) the computational approach (Section 5 with reference to [8,22]), c) the simplified computational method (Section 5 and also with reference to [18,22]), capture quite well the two different behaviours that are carried out along the directions $x_1$ and $x_2$. Namely, these models describe quite accurately the behaviour characterised by great characteristic lengths along the layering (Figure 4.a) and by vanishing characteristic lengths along the normal direction to the layers (Figures 4.c and 4.d).

To further evaluate this result, let us now consider the shear deformation of a two-dimensional strip made up of the three-phase composite. The strip has height $L = 10\varepsilon$ and is unlimited in the horizontal direction. The layers in this laminated composite are considered horizontal (horizontal-layer model in [18]). In the heterogeneous model the lower edge is considered restrained, while a horizontal displacement $\Delta = L/100$ is prescribed to the upper edge (see [18,22] for details and for the analytical solution of the macroscopic problem in terms of the characteristic lengths $\lambda_{Sh-2}$). In Figure 5 is shown the component $H_{12}$ of the macro-displacement gradient in terms of the ratio $\tilde{x}_2/L$ for horizontal-layer model with $\eta = E_3/E_2 = 100$, $\psi = 2000/11$ and $\nu_1 = \nu_2 = \nu_3 = 0.1$. From this diagram, a good agreement is observed between the results obtained by the second-order model, characterized by the asymptotic homogenization approach (red line) and by simplified computational homogenization method (green line) (red and green lines are coincident), and the heterogeneous model (diamonds). Conversely, the computational homogenization approach (blue line) provides a lower approximation of the result obtained by the heterogeneous model.

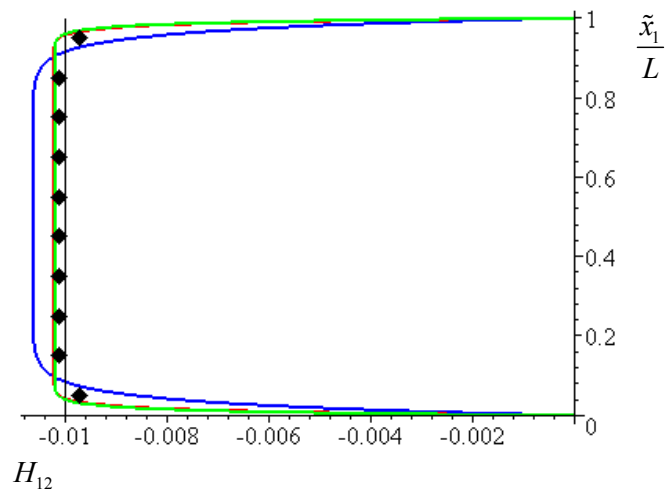

Figure 5: Boundary shear layer problem: the component $H_{12}$ of the macro-displacement gradient. Heterogeneous model (diamonds) versus homogeneous second-order and first-order models (Second order model: red line asymptotic approach; blue line computational approach; green line simplified computational approach [18]. First order model black line).

Conversely, for extensional problems along axes $x_1$ (Figure 4.b), the results obtained from homogenized models differ considerably from those obtained from the heterogeneous model. The macro-displacement of the second-order model, with overall elastic moduli evaluated through the asymptotic and simplified computational approaches, is approximately equal to that obtained from the first-order model (i.e. red, green and black lines are coincident) while the heterogeneous model returns macro-displacements that are considerably larger. On the other hand, the difference between the second-order model characterised through the computational approach (blue line) and the heterogeneous model is considerably greater. The homogenization model proposed in [18], although being a simplified approach, provides result in agreement with those of the asymptotic approach and is less computationally burdensome than the second one.

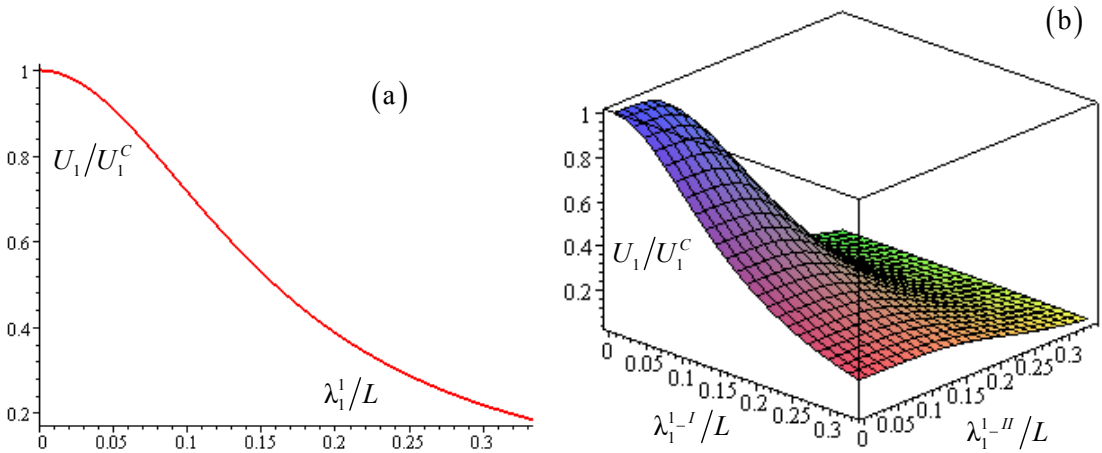

Figure 6: Dimensionless macro-displacement $U_1/U_1^C$ versus characteristic lengths: (a) second-order model, (b) third-order model.

It may be verified that, for shear and extensional problem, when the displacement field in the heterogeneous model is greater than that obtained through a first-order homogeneous model, a second gradient continuum (or a multipolar continuum) turns out to be unsuitable model. In fact, the macro-displacement obtained by this continuum model cannot be greater than the displacement $U_\beta^C$ evaluated by a first-order homogeneous model as shown in Figure 6 and by equations (44) and (45) in Appendix D. This result is due to the structure of the equilibrium equation (44) in terms of the macro-displacement $U_\beta$ while it does not depend on the characteristic length $\lambda \in \mathbb{R}$ of the non-local continuum determined by a homogenization techniques (see equations (45) and (46)). It should be noted that, in order to obtain $U_\beta \geq U_\beta^C$ a not realistic characteristic length $\lambda \in \mathbb{C}$ is required. Notably, in

Figure 6.a the dimensionless displacement $U_1/U_1^C$ at a point of the body (see equation (45)) is shown in terms of the ratio $\lambda_1^1/L$ (the solution of the non-local homogeneous problem is shown in Appendix D). Similar results are shown in Figure 6.b where the displacement ratio $U_1/U_1^C$ is obtained by a third-order homogeneous continuum for varying the two dimensionless characteristic lengths $\lambda_1^{1-I}/L$, $\lambda_1^{1-II}/L$. The dimensionless macro-displacement $U_\beta/U_\beta^C$ obtained in case of extensional problem along the direction $x_2$ or of shear problems along the directions $x_1$, $x_2$ is similar to that shown in the diagrams of Figure 6 for $U_1/U_1^C$. These final outcomes may suggest a question: may better results be achieved by homogenizing the periodic material in a homogeneous micromorphic continuum (or generalised micromorphic continuum) that has additional degrees of freedom? Nevertheless, it is worth to note that the characterisation of the elastic moduli of such continua result rather complex and many aspects have to be clarified [19].

## 7 Conclusions

The second-order asymptotic and computational homogenization techniques here proposed provides continuous and sufficiently regular micro-displacement fields that ensure the anti-periodicity of traction on the boundary of the periodic cell and satisfies the weak form of the local equilibrium equations. The perturbation functions to be included in the description of the micro-displacement field are obtained as the solution of non-homogeneous elasticity problems on the periodic cell with periodic boundary conditions and normalization condition. The overall elastic moduli derived by the macro-homogeneity condition are not affected by the choice of periodic cell. In the limit case of locally homogeneous material, i.e. of an evanescent microstructure, the characteristic lengths vanish, a circumstance that highlights the absence of non-local effects at the macroscopic scale. The approximation of the macro-displacement field through a Taylor polynomial expansion (with the objective of establishing correlations between the asymptotic and computational homogenization approaches) leads to a simplification in the evaluation of the overall elastic moduli in which the contribution associated to the tensor $A^{H-\kappa}_{p_1q_1r_1p_2q_2r_2}$ that is related to the third-order strain is neglected. This approximation of the macro-displacement field and the further assumption of vanishing body forces makes the homogenization procedure less computationally cumbersome (see [18,22]). In this case the perturbation functions are determined by the solution of elasticity problems on the periodic cell with generalised boundary conditions and the local equilibrium is satisfied inside the cell but the micro-displacement obtained for localization does not result sufficiently regular to ensure the anti-periodicity of the traction on the boundary of the periodic cell. In the developed examples the differences between a) the homogenization model based on an asymptotic expansion of the mean strain energy elaborated here, b) the computational approach (that provides overall elastic moduli that have the same shape of those obtained through different procedures in [22] and in [8] in the case of strain gradient continuum) and c) the simplified computational homogenization method [18,22], are highlighted. The characteristic lengths obtained by these methods are greater than those obtained through the more accurate asymptotic approach independently of the ratio between the elastic moduli of the constituents. Notably, these differences result more significant in cases where the characteristic lengths (evaluated through the more accurate asymptotic approach) are small. The displacement obtained by both the asymptotic approach and the simplified computational approach is greater than that obtained by the computational approach but all these displacements are lower than the one returned by a first-order homogeneous continuum. The simplified computational homogenization developed in [18,22] provides result in good agreement with those of the asymptotic approach. Finally, from the considered examples it follows that the second order model and the multipolar models do not seem to be suitable to describe the behaviour of heterogeneous materials when the displacement obtained by the heterogeneous model is greater than that determined through a first-order homogeneous model. In these cases, a question arises concerning the possibility to achieve better results by homogenizing the periodic material in a homogeneous micromorphic continuum (or generalised micromorphic continuum) where additional degrees of freedom with respect to the displacement are embedded. Nevertheless, it is worth to note that the characterisation of the elastic moduli of such continua result rather complex and many aspects have to be clarified.

**Acknowledgment**

The author acknowledges financial support from (MURST) Italian Department for University and Scientific and Technological Research under the framework of the research MIUR Prin09 project 2009XWLFKW, Multi-scale problems with complex interactions in Structural Engineering, coordinated by prof. A. Corigliano.

**Appendix A**

The line integral along the boundary $\partial \mathscr{L}$ in the mean strain energy expressed in equation (16) is zero because of the $\mathscr{L}$-periodicity of the macro and of the macro-displacement strain tensor. Therefore, the mean strain energy takes the following form

$$\mathscr{E}_m = \frac{1}{2}\left\langle C^m_{ijkl}\tilde{B}^H_{ijpq}\tilde{B}^H_{klp_1q_1}\right\rangle \int_{\mathscr{L}} H_{pq}H_{p_1q_1}d\mathbf{x} + \varepsilon \left\langle C^m_{ijkl}\tilde{B}^H_{ijpq}\tilde{B}^\kappa_{klp_1q_1r_1}\right\rangle \int_{\mathscr{L}} H_{pq}\kappa_{p_1q_1r_1}d\mathbf{x} +$$
$$+ \frac{\varepsilon^2}{2}\Bigg\{\left\langle C^m_{ijkl}\tilde{B}^\kappa_{ijpqr}\tilde{B}^\kappa_{klp_1q_1r_1}\right\rangle \int_{\mathscr{L}} \kappa_{pqr}\kappa_{p_1q_1r_1}d\mathbf{x} - \frac{1}{3}\Big\langle C^m_{stir_2}\tilde{A}^\kappa_{ip_1q_1r_1}\tilde{B}^H_{stp_2q_2} + C^m_{sr_2it}\tilde{B}^H_{itp_2q_2}\tilde{A}^\kappa_{sp_1q_1r_1} +$$
$$+ C^m_{stir_1}\tilde{A}^\kappa_{ip_1q_1r_2}\tilde{B}^H_{stp_2q_2} + C^m_{sr_1it}\tilde{B}^H_{itp_2q_2}\tilde{A}^\kappa_{sp_1q_1r_2} + C^m_{stir_1}\tilde{A}^\kappa_{ip_1r_2q_1}\tilde{B}^H_{stp_2q_2} +$$
$$+ C^m_{sr_1it}\tilde{B}^H_{itp_2q_2}\tilde{A}^\kappa_{sp_1r_2q_1}\Big\rangle \int_{\mathscr{L}} \kappa_{p_1q_1r_1}\kappa_{p_2q_2r_2}d\mathbf{x}\Bigg\} \quad (33)$$

and in the more synthetic form given in equation (17). Notably, the localization tensors $B^H_{ijpq}$, $B^\kappa_{ijpqr}$ and $A^{H\_\kappa}_{p_1q_1r_1p_2q_2r_2}$ in the *mean strain energy* (17) are obtained by symmetrization with respect to the repeated indices of the terms in equation (16) and are expressed in the form

$$B^H_{ijpq}\left(\boldsymbol{\xi} = \frac{\mathbf{x}}{\varepsilon}\right) = \frac{1}{4}\left(\delta_{ip}\delta_{jq} + N^1_{ipq,j} + \delta_{jp}\delta_{iq} + N^1_{jpq,i} + \delta_{iq}\delta_{jp} + N^1_{iqp,j} + \delta_{jq}\delta_{ip} + N^1_{jqp,i}\right),$$

$$B^\kappa_{ijpqr}\left(\boldsymbol{\xi} = \frac{\mathbf{x}}{\varepsilon}\right) = \frac{1}{4}\left(N^1_{ipq}\delta_{jr} + N^1_{ipr}\delta_{qj} + 2N^2_{ipqr,j} + N^1_{jpq}\delta_{ir} + N^1_{jpr}\delta_{qi} + 2N^2_{jpqr,i}\right),$$

$$A^{H\_\kappa}_{p_1q_1r_1p_2q_2r_2}\left(\boldsymbol{\xi} = \frac{\mathbf{x}}{\varepsilon}\right) = \Big(C^m_{p_2q_2ir_2}N^2_{ip_1q_1r_1} + C^m_{p_2q_2ir_1}N^2_{ip_1q_1r_2} + C^m_{p_2q_2iq_1}N^2_{ip_1r_1r_2} + C^m_{p_2r_2iq_2}N^2_{ip_1q_1r_1} +$$
$$+ C^m_{p_2r_2ir_1}N^2_{ip_1q_1q_2} + C^m_{p_2r_2iq_1}N^2_{ip_1r_1q_2} + C^m_{p_1q_1ir_1}N^2_{ip_2q_2r_2} + C^m_{p_1q_1ir_2}N^2_{ip_2q_2r_1} + C^m_{p_1q_1iq_2}N^2_{ip_2r_2r_1} +$$
$$+ C^m_{p_1r_1iq_1}N^2_{ip_2q_2r_2} + C^m_{p_1r_1ir_2}N^2_{ip_2q_2q_1} + C^m_{p_1r_1iq_2}N^2_{ip_2r_2q_1}\Big) + \frac{1}{2}\Big(C^m_{stir_1}N^2_{ip_2q_2r_2}N^1_{sp_1q_1,t} + C^m_{sr_1it}N^2_{sp_2q_2r_2}N^1_{ip_1q_1,t} +$$
$$+ C^m_{sr_2it}N^2_{sp_2q_2r_1}N^1_{ip_1q_1,t} + C^m_{stiq_2}N^2_{ip_2r_2r_1}N^1_{sp_1q_1,t} + C^m_{sq_2it}N^2_{sp_2r_2r_1}N^1_{ip_1q_1,t} + C^m_{stiq_1}N^2_{ip_2q_2r_2}N^1_{sp_1r_1,t} + \quad (34)$$
$$+ C^m_{sq_1it}N^2_{sp_2q_2r_2}N^1_{ip_1r_1,t} + C^m_{stir_2}N^2_{ip_2q_2q_1}N^1_{sp_1r_1,t} + C^m_{sr_2it}N^2_{sp_2q_2q_1}N^1_{ip_1r_1,t} + C^m_{stiq_2}N^2_{ip_2r_2q_1}N^1_{sp_1r_1,t} +$$
$$+ C^m_{sq_2it}N^2_{sp_2r_2q_1}N^1_{ip_1r_1,t} + C^m_{stir_2}N^2_{ip_1q_1r_1}N^1_{sp_2q_2,t} + C^m_{sr_2it}N^2_{sp_1q_1r_1}N^1_{ip_2q_2,t} + C^m_{stir_1}N^2_{ip_1q_1r_2}N^1_{sp_2q_2,t} +$$
$$+ C^m_{sr_1it}N^2_{sp_1q_1r_2}N^1_{ip_2q_2,t} + C^m_{stiq_1}N^2_{ip_1r_1r_2}N^1_{sp_2q_2,t} + C^m_{sq_1it}N^2_{sp_1r_1r_2}N^1_{ip_2q_2,t} + C^m_{stiq_2}N^2_{ip_1q_1r_1}N^1_{sp_2r_2,t} +$$
$$+ C^m_{sq_2it}N^2_{sp_1q_1r_1}N^1_{ip_2r_2,t} + C^m_{stir_1}N^2_{ip_1q_1q_2}N^1_{sp_2r_2,t} + C^m_{sr_1it}N^2_{sp_1q_1q_2}N^1_{ip_2r_2,t} + C^m_{stiq_1}N^2_{ip_1r_1q_2}N^1_{sp_2r_2,t} +$$
$$+ C^m_{sq_1it}N^2_{sp_1r_1q_2}N^1_{ip_2r_2,t} + C^m_{stir_2}N^2_{ip_2q_2r_1}N^1_{sp_1q_1,t}\Big).$$

**Appendix B**

The homogenization approach proposed in Section 4 may be specialised to the case homogenization of a Koiter continuum. According to [8] the second-order strain $\kappa_{ijp}$ may be expressed in term of the curvature $\kappa_{\alpha r}$ by the following relation

$$\kappa_{ijp} = -\varepsilon_{ij\alpha}\kappa_{\alpha p} \qquad (i,j,p = 1,2 \text{ e } \alpha = 3) \qquad (35)$$

that allows to express, recalling (9), the micro-displacement field $\hat{\mathbf{u}}^{II}(\mathbf{x})$ in the form

$$\hat{u}_i^{II}(\mathbf{x}) = \hat{u}_i^{II}\left(\mathbf{x}, \boldsymbol{\xi} = \frac{\mathbf{x}}{\varepsilon}\right) = \left(U_i(\mathbf{x}) + \varepsilon N_{ipq}^1(\boldsymbol{\xi}) H_{pq}(\mathbf{x}) - \varepsilon^2 N_{ipqr}^2(\boldsymbol{\xi}) \varepsilon_{pq\alpha} \kappa_{\alpha r}(\mathbf{x})\right)\bigg|_{\boldsymbol{\xi} = \frac{\mathbf{x}}{\varepsilon}}, \qquad (36)$$

where $\varepsilon_{pq\alpha}$ is the permutation symbol.

In analogy with the procedure proposed in Section 4 for the second-order continuum (see equations (10)-(17)) the mean strain energy $\mathcal{E}_m$ may be determined in terms of localization tensors $B_{ijpq}^H$, $B_{kl\alpha r_1}^\kappa$, $A_{\alpha r_1 \beta r_2}^{H\_\kappa}$, i.e.

$$\mathcal{E}_m = \frac{1}{2}\left\langle C_{ijkl}^m B_{ijpq}^H B_{klp_1q_1}^H \right\rangle \int_\mathscr{L} H_{pq} H_{p_1q_1} d\mathbf{x} + \varepsilon \left\langle C_{ijkl}^m B_{ijpq}^H B_{kl\alpha r_1}^\kappa \right\rangle \int_\mathscr{L} H_{pq} \kappa_{\alpha r_1} d\mathbf{x} + \\ + \frac{\varepsilon^2}{2}\left\langle C_{ijkl}^m B_{ij\alpha r}^\kappa B_{kl\beta r_1}^\kappa \right\rangle \int_\mathscr{L} \kappa_{\alpha r} \kappa_{\beta r_1} d\mathbf{x} - \frac{\varepsilon^2}{24}\left\langle A_{\alpha r_1 \beta r_2}^{H\_\kappa} \right\rangle \int_\mathscr{L} \kappa_{\alpha r_1} \kappa_{\beta r_2} d\mathbf{x}, \qquad (37)$$

where $B_{ij\alpha r}^\kappa = -\varepsilon_{pq\alpha} B_{ijpqr}^\kappa$ and $A_{\alpha r_1 \beta r_2}^\kappa = \varepsilon_{p_1q_1\alpha}\varepsilon_{p_2q_2\beta}A_{p_1q_1r_1p_2q_2r_2}^{H\_\kappa}$ with $B_{ijpqr}^\kappa$ and $A_{p_1q_1r_1p_2q_2r_2}^{H\_\kappa}$ expressed in equation (34). Through the generalised macro-homogeneity condition, between macroscopic strain energy

$$\mathcal{E}_M = \frac{1}{2}C_{ijkl}\int_\mathscr{L} H_{ij}H_{kl}d\mathbf{x} + \hat{Y}_{ij\alpha p}\int_\mathscr{L} H_{ij}\kappa_{\alpha p}d\mathbf{x} + \frac{1}{2}\hat{S}_{\alpha p \beta j}\int_\mathscr{L} \kappa_{\alpha r}\kappa_{\beta p}d\mathbf{x}, \qquad (\alpha, \beta = 3) \qquad (38)$$

and the mean strain energy (see equation (37)), the elastic moduli of the Koiter continuum are determined, and take the form

$$C_{pqp_1q_1} = \left\langle C_{ijkl}^m B_{ijpq}^H B_{klp_1q_1}^H \right\rangle, \quad \frac{\hat{Y}_{pq\alpha r}}{\varepsilon} = \left\langle C_{ijkl}^m B_{ijpq}^H B_{kl\alpha r}^\kappa \right\rangle, \quad \frac{\hat{S}_{\alpha p \beta j}}{\varepsilon^2} = \left\langle C_{ijkl}^m B_{ij\alpha p}^\kappa B_{kl\beta r}^\kappa \right\rangle - \frac{1}{12}\left\langle A_{\alpha p \beta j}^{H\_\kappa} \right\rangle, \quad (39)$$

that coincide with (19) apart from suitable index contractions, associated with definition (35), of the higher-order elastic moduli.

Computational homogenization of a Koiter continuum is obtained specialising the model proposed in Section 5. Through (26) and (35) the micro-displacement $\hat{\mathbf{u}}^{II}(\mathbf{x})$ expressed in (36) takes the form

$$\hat{u}_i^{II}(\mathbf{x}) = \hat{u}_i^{II}\left(\mathbf{x}, \boldsymbol{\xi} = \frac{\mathbf{x}}{\varepsilon}\right) = \bar{U}_i + \bar{H}_{iq} x_q - \frac{1}{2} \varepsilon_{iq\alpha} \bar{\kappa}_{\alpha r} x_q x_r +$$
$$-\left(\frac{\varepsilon}{2}\left[N^1_{ipq}(\boldsymbol{\xi}) x_r + N^1_{ipr}(\boldsymbol{\xi}) x_q\right] \varepsilon_{pq\alpha} \bar{\kappa}_{\alpha r} - \varepsilon N^1_{ipq}(\boldsymbol{\xi}) \bar{H}_{pq} + \varepsilon^2 N^2_{ipqr}(\boldsymbol{\xi}) \varepsilon_{pq\alpha} \bar{\kappa}_{\alpha r}\right)\bigg|_{\boldsymbol{\xi}=\frac{\mathbf{x}}{\varepsilon}}, \quad (40)$$

where $\bar{\kappa}_{\alpha r}$ are the curvature evaluated in the barycentre of the periodic cell, i.e. $\bar{\kappa}_{\alpha r} = \kappa_{\alpha r}(\mathbf{x}=\mathbf{0})$. The elastic moduli of continuous homogeneous Koiter are determined through the generalised macro-homogeneity condition such as in the asymptotic approach here developed earlier in this appendix, and take the form

$$C_{pqp_1q_1} = \left\langle C^m_{ijkl} B^H_{ijpq} B^H_{klp_1q_1} \right\rangle, \quad \frac{\hat{Y}_{pq\alpha r}}{\varepsilon} = \left\langle C^m_{ijkl} B^H_{ijpq} B^\kappa_{kl\alpha r} \right\rangle, \quad \frac{\hat{S}_{\alpha p\beta j}}{\varepsilon^2} = \left\langle C^m_{ijkl} B^\kappa_{ij\alpha p} B^\kappa_{kl\beta r} \right\rangle. \quad (41)$$

**Appendix C**

The computational homogenization proposed in Section 5 may be generalised to the case of multi-polar homogenization if the micro-displacement $\mathbf{u}^N(\mathbf{x})$ is expressed as super position of $N+1$ contributions, i.e.

$$u_i^N(\mathbf{x}) = u_i^N\left(\mathbf{x}, \boldsymbol{\xi} = \frac{\mathbf{x}}{\varepsilon}\right) = U_i(\mathbf{x}) + \sum_{j=1}^{N} \varepsilon^j \sum_{t=0}^{N-1} \frac{1}{t!} \sum_{|q|=j+1} \sum_{|p|=t} N^j_{iq}(\boldsymbol{\xi}) U^j_{qp} \mathbf{x}^p, \quad (42)$$

where the macro-displacement $U_i(\mathbf{x})$ is assumed in the form

$$U_i(\mathbf{x}) = U_i^0 + \sum_{j=1}^{N} \sum_{|q|=j} U^j_{iq} \mathbf{x}^q = U_i^0 + U^1_{iq_1} x_{q_1} + U^2_{iq_1q_2} x_{q_1} x_{q_2} + \ldots + U^N_{iq_1q_2\ldots q_N} x_{q_1} x_{q_2} \ldots x_{q_N},$$

and being $U_i^0 = U_i(\mathbf{x}=\mathbf{0})$, $U^1_{iq_1} = H_{iq_1}(\mathbf{x}=\mathbf{0}) = \bar{H}_{iq_1}$, $U^2_{iq_1q_2} = \kappa_{iq_1q_2}(\mathbf{x}=\mathbf{0}) = \bar{\kappa}_{iq_1q_2}$ and $U^N_{iq_1q_2\ldots q_N} = \kappa_{pq_1q_2\ldots q_N}(\mathbf{x}=\mathbf{0})$. The perturbation functions $N^m_{ipq_1\ldots q_m}$ are determined through the solution of the first $N$ cell problems (equation (6)). Therefore, the micro-displacement gradient may be expressed in the form

$$\frac{\partial u_i^N(\mathbf{x})}{\partial x_j} = \tilde{B}_{ijpq_1}^H\left(\mathbf{x}, \boldsymbol{\xi}=\frac{\mathbf{x}}{\varepsilon}\right)H_{pq_1} + \varepsilon \tilde{B}_{ijpq_1q_2}^K\left(\mathbf{x}, \boldsymbol{\xi}=\frac{\mathbf{x}}{\varepsilon}\right)\kappa_{pq_1q_2} + \ldots\ldots +$$
$$+ \varepsilon^{N-1}\tilde{B}_{ijpq_1q_2\ldots q_N}^K\left(\mathbf{x}, \boldsymbol{\xi}=\frac{\mathbf{x}}{\varepsilon}\right)\kappa_{pq_1q_2\ldots q_N}, \tag{43}$$

where $\tilde{B}_{ijpq}^H$, $\tilde{B}_{ijpq_1q_2}^K$ and $\tilde{B}_{ijpq_1q_2\ldots q_N}^K$ are suitable localization tensors which depends of the perturbation functions. The elastic moduli of the homogeneous multi-polar continuum are determined through the generalised macro-homogeneity condition, in terms of localization tensors $B_{ijpq}^H$, $B_{ijpq_1q_2}^K$ and $B_{ijpq_1q_2\ldots q_N}^K$ suitably symmetrised with respect to the saturated indices.

**Appendix D**

With reference to Section 6, the equilibrium of an elastic body with applied body forces $f_\beta$ $(\beta=1,2)$ is described through a second-order continuum and may be analysed through a displacement formulation. In the case to two-dimensional bodies having a microstructure characterised by periodic cell with orthogonal symmetry axis $x_\alpha$ $(\alpha=1,2)$ and body forces applied in the direction of this axis, the equilibrium equations take the form

$$-S_{\beta\alpha\alpha\beta\alpha\alpha}U_{\beta,\alpha\alpha\alpha\alpha} + C_{\beta\alpha\beta\alpha}U_{\beta,\alpha\alpha} = -f_\beta, \tag{44}$$

(with indices $\alpha,\beta$ not summed) that describes both the shear problem $(\alpha \neq \beta)$ and extensional problem $(\alpha = \beta)$. For the harmonic body forces $f_\beta(x_\alpha) = \Xi_\beta e^{i\left(\frac{2\pi}{L_\alpha}x_\alpha\right)}$ $(\Xi_\beta \in \mathbb{R})$ the macro-displacement is

$$U_\beta(x_\alpha) = \left(\frac{L_\alpha}{2\pi}\right)^2 \frac{\Xi_\beta e^{i\left(\frac{2\pi}{L_\alpha}x_\alpha\right)}}{C_{\beta\alpha\beta\alpha}\left[1 + \left(2\pi\frac{\lambda_\alpha^\beta}{L_\alpha}\right)^2\right]}, \tag{45}$$

from which one observe that if the characteristic lengths $\lambda_\alpha^\beta$ is vanishing the displacement $U_\beta^C = \left(\frac{L_\alpha}{2\pi}\right)^2 \Xi_\beta e^{i\left(\frac{2\pi}{L_\alpha}x_\alpha\right)} \Big/ C_{\beta\alpha\beta\alpha}$ of a first-order homogeneous continuum is obtained. For multi-polar continua of order $N$ the relation (45) may be generalised, for the ellipticity of the equilibrium equations [2,9], in terms of $N-1$ characteristic lengths $\lambda_\alpha^{\beta_{-j}}$ and results

$$U_\beta(x_\alpha) = \left(\frac{L_\alpha}{2\pi}\right)^2 \frac{\Xi_\beta \, e^{i\left(\frac{2\pi}{L_\alpha}x_\alpha\right)}}{C_{\beta\alpha\beta\alpha} \sum_{j=1}^{N-1}\left[1+\left(2\pi\frac{\lambda_\alpha^{\beta_{-j}}}{L_\alpha}\right)^{2j}\right]}. \qquad (46)$$